\documentclass[12pt]{iopart}

\ifx\pdfoutput\undefined
   \usepackage[dvips]{graphicx}
   \else
   \usepackage[pdftex]{graphicx}
   \fi
   \usepackage{epstopdf}

\usepackage{graphics}
\usepackage{graphicx}
\usepackage{epsfig}
\begin{document}

\DeclareGraphicsExtensions{.pdf,.gif,.jpg}

\title[BayesWave]{BayesWave: Bayesian Inference for Gravitational Wave Bursts and Instrument Glitches}

\author{Neil J. Cornish}

\address{Department of Physics, Montana State University, Bozeman,
MT 59717, USA}

\author{Tyson B. Littenberg}

\address{Center for Interdisciplinary Exploration and Research in
Astrophysics (CIERA) \& Department of Physics and Astronomy, Northwestern University,
2145 Sheridan Road, Evanston, IL 60208}

\begin{abstract}
A central challenge in Gravitational Wave Astronomy is identifying weak signals in the presence of
non-stationary and non-Gaussian noise. The separation of gravitational wave signals from noise requires
good models for both. When accurate signal models are available, such as
for binary Neutron star systems, it is possible to make robust detection statements even when the noise
is poorly understood. In contrast, searches for ``un-modeled'' transient signals are strongly impacted by the
methods used to characterize the noise. Here we take a Bayesian approach and introduce a multi-component,
variable dimension, parameterized noise model  that explicitly accounts for non-stationarity and non-Gaussianity
in data from interferometric gravitational wave detectors. Instrumental transients (glitches) and burst sources
of gravitational waves are modeled using a Morlet-Gabor continuous wavelet frame. The number and placement of
the wavelets is determined by a trans-dimensional Reversible Jump Markov Chain Monte Carlo algorithm.
The Gaussian component of the noise and sharp line features in the noise spectrum are modeled
using the {\em BayesLine} algorithm, which operates in concert with the wavelet model.
\end{abstract}



\maketitle

\section{Introduction}

The LIGO~\cite{ligo} and Virgo~\cite{virgo} gravitational wave detectors are currently undergoing major upgrades,
with the goal of improving the broadband sensitivity by an order of magnitude~\cite{TheLIGOScientific:2014jea,TheVirgo:2014hva}.
But it takes more than lasers, mirrors, photodiodes and suspension systems to detect gravitational waves. Another
key component is the signal processing used to lift the faint gravitational wave signals out of the instrument noise. 

Searches for gravitational waves using data from the initial LIGO-Virgo science runs have shown that non-Gaussian
features in the data, such as noise transients or ``glitches'', impact the ability to detect weak signals as the various detection
statistics used in the searches end up with heavier tails than they would for Gaussian noise~\cite{Blackburn:2008ah,Abbott:2007kv,Aasi:2012wd}.
The standard procedure for accounting for these effects is to perform Monte Carlo trials with scrambled data produced by introducing time lags 
that break signal coherence between the detectors in the network. These trials are used to
produce background probability distributions for search statistics -- such as the cross correlation of signal templates with the
data -- which can then be used to assess the significance of events detected in the zero-lag data. Extensive testing with simulated
signal injections into the data has shown that this standard approach is reliable and robust, though
with a sensitivity that is less than what would be possible if the noise were stationary and Gaussian. The time-slide approach
also has limitations in terms of the detection significance that can be reached~\cite{Was:2010zz}. In addition to degrading
detection efficiency, the non-Gaussianity of the noise can also impact parameter estimation of quantities like the sky location
and masses of a binary system~\cite{Aasi:2013jjl}.

The detection statistics used in the searches are often motivated by analytical studies assuming Gaussian noise, and then
the effects of non-Gaussianity are taken into account by using the Monte Carlo derived distributions estimated from the
actual detector data. In effect the data are used to infer the noise properties, or at least, the properties of the noise
projected onto the signal manifold defined by the detection statistic. Here the noise modeling is taken a step further in
a comprehensive approach to gravitational wave detection that employs a multi-component parameterized noise model in addition to the
signal model.  The data are used to jointly infer the parameters of both the noise and signal models, following the motto
``model everything and let the data sort it out"\footnote{With apologies to the two excellent referees for our paper who disliked this
motto and requested that we remove it.}. Of course, it is not possible to model every possible feature in the data, but
our ultimate goal is to develop detailed models that can account for a wide range of non-stationary and non-Gaussian
contributions to the instrument noise. The algorithm described here is a first step towards that goal.
A key feature of the noise and signal models is that their degree of complexity is not fixed in advance,
but rather, determined by Bayesian model selection. For example, short duration, non-stationary noise transients are modeled using
a sum of continuous wavelets, but the number of wavelets is not fixed. Stretches of data with loud glitches that are spread over
a wide time-frequency volume will be described by a high dimensional model, while quiet stretches of data will be described by
a low dimensional model. Bayesian model selection naturally finds the appropriate balance between model fidelity and model
complexity, so there is no danger of ``over fitting'' the data.

In an earlier study we introduced a discrete orthogonal wavelet model to account for short-duration, non-stationary features in the instrument
noise~\cite{Littenberg:2010gf}. This approach was developed further in collaboration with P. Baker, and a description of the study can
be found in Baker's Ph.D. thesis~\cite{pbaker}. From these foundations we have developed the {\em BayesWave} algorithm for the joint detection and
characterization of instrument glitches and gravitational wave bursts. The types of signals {\em BayesWave} targets include core collapse supernovae, stellar
mass black hole mergers and unexpected short duration signals (up to a few seconds in duration). {\em BayesWave} in its current form is not well suited for
studying longer duration signals such as neutron star binary inspirals, isolated deformed neutron stars or stochastic backgrounds.

The name ``{\em BayesWave}'' is derived from a concatenation of ``Bayesian'' and ``Wavelets.'' The key to the success of the new algorithm is the adoption of
a non-orthogonal, continuous Morlet-Gabor wavelet frame\footnote{A frame generalizes the concept of a basis to allow for sets that are linearly dependent}, which
provides an extremely compact representation of bursts and glitches,
and has the additional advantage that the entire analysis can be implemented in the frequency domain.  Bayesian model selection and
parameter inference is implemented using a trans-dimensional
Reversible Jump Markov Chain Monte Carlo (RJMCMC)~\cite{Green} algorithm that varies both the model parameters and the model dimension (in our case the
number and parameters of the wavelets used to describe the
glitches and gravitational wave signals).  Using a variable number of wavelets builds in a natural parsimony that balances the complexity of the model against the quality of the fit.
Much of the {\em BayesWave} algorithm has been re-purposed from techniques we developed to detect and characterize thousands of overlapping signals
from compact white dwarf binaries in simulated data from a future space based gravitational wave detector~\cite{Crowder:2006eu, Littenberg:2011zg}.
The {\em BayesWave} algorithm is run in concert with the {\em BayesLine} spectral whitening algorithm, which we describe in a companion paper~\cite{Littenberg:2014oda}.
The {\em BayesLine} algorithm models the noise spectrum using a  combination of a smooth spline curve and a collection of
Lorentzians to fit sharp line features. {\em BayesLine} employs a RJMCMC algorithm to vary the number and placement of the
spline control points and the number of Lorentzians and their parameters.

We begin by reviewing the application of Bayesian inference in gravitational wave astronomy, with an emphasis on the importance of choosing
an appropriate likelihood function. Next we describe the glitch and gravitational wave burst models and lay out our reasons for choosing
the Morlet-Gabor wavelet frame. This is followed by an investigation of various time-frequency clustering priors using simulated Gaussian noise and
simulated gravitational wave signals. The effectiveness of the {\em BayesWave} algorithm is then demonstrated using real-world data from the
sixth LIGO science run, including glitch detection and reconstruction, and the recovery of hardware and software signal injections.

\section{Bayesian Inference for Gravitational Wave Astronomy}

The posterior distribution function $p({\bf h} \vert {\bf s}, M)$ summarizes the properties of a gravitational waveform ${\bf h}$ that can be inferred from data ${\bf s}$ under model $M$:
\begin{equation}
p({\bf h} \vert {\bf s}, M) = \frac{p({\bf h} \vert M) p({\bf s}\vert {\bf h}, M)}{p({\bf s}\vert M)} \, ,
\end{equation}
where $p({\bf h}\vert M)$ is the prior distribution for ${\bf h}$ in model $M$,
$p({\bf s}\vert {\bf h},M)$ is the likelihood of the data for the given waveform, and 
$p({\bf s} \vert M) = \int p({\bf h} \vert M) p({\bf s}\vert {\bf h},M) d{\bf h}$
is the marginal likelihood, or evidence, for the model $M$ under consideration. Here
${\bf s}$ is a vector of data collected from a network of gravitational wave detectors,
which is a combination of instrument noise ${\bf n}$ and a gravitational
wave signal ${\bf h}$ such that ${\bf s} = {\bf R}\star {\bf h} + {\bf n}$, where ${\bf R}$ is a time-delay operator that describes the network response to the two gravitational
wave polarizations ${\bf h}$.

The Bayesian approach to data analysis is entirely mechanical. Once you have specified your signal model and the likelihood function, the posterior distribution
can be computed using numerical techniques such as Markov Chain Monte Carlo~\cite{gg} or Nested Sampling~\cite{skilling}. From the
posterior distribution function we can extract point estimates and ``error bars'' for the parameters that
characterize the waveform ${\bf h}$, and by comparing the evidence for alternative models, we can
quote odds ratios for possibilities such as ``the data contains the signal from a black
hole merger'' and ``the data is purely instrumental noise.'' On the other hand, the quality of what you get
out is only as good as what you put in. If your signal model is flawed, or your likelihood function is too
simple, the posterior distribution will produce misleading results. The frequentist approach is more forgiving,
as the likelihood function is only used to motivate the form of a search statistic, which is then tuned
using signal injections and time slides of the data (see e.g. Ref.~\cite{Abbott:2007xi}).

The likelihood function is defined by demanding that the residual formed from subtracting the gravitational wave model from the
data, ${\bf r} = {\bf s} - {\bf R}\star {\bf h}$, is consistent with the model for the noise ${\bf n}$.
For stationary, Gaussian noise this leads to the standard likelihood function~\cite{Finn:1998vp} 
\begin{equation}\label{like}
p({\bf s}\vert {\bf h}, {\cal M}) = \frac{ \exp\left[-({\bf r}^T \cdot {\bf C}^{-1}\cdot {\bf r} )/2 \right]}
{((2\pi)^N {\rm det} {\bf C})^{1/2}}
\end{equation}
where ${\bf C}$ is the noise correlation matrix and $N$ is the length of the data ${\bf s}$. For stationary noise, transforming the data
to the frequency domain yields a diagonal correlation matrix, $C_{f \, f'} = S_n(f) \, \delta_{f\, f'} $, which allows for faster likelihood evaluations.
The existing Bayesian parameter estimate algorithms that have been applied to LIGO/Virgo data~\cite{Aasi:2013jjl} assume that the noise is stationary and
Gaussian and use off-source estimates for the noise spectrum $S_n(f)$. However, non-stationary, non-Gaussian features such as glitches or wandering lines can bias this
type of analysis~\cite{Littenberg:2010gf,Littenberg:2013gja}.

The need for more realistic noise modeling has been emphasized in several studies.
Allen {\it et al.}~\cite{Allen:2002jw,Allen:2001ay} have argued that parameterized noise models are
needed to derive robust search statistics. They found that distributions with heavier tails than a Gaussian
were more robust against noise transients, a result that has been confirmed by R\"over using a Student-t likelihood function~\cite{Rover:2011qd}.
Clark {\it et al.}~\cite{Clark:2007xw} included a model
for Sine-Gaussian instrument glitches as an alternative hypothesis when computing Bayesian odds ratios
for gravitational wave signals from pulsar glitches.  Clark {\it et al.} also suggested that it would be
valuable to have a classification of instrument glitches that could be used to construct better
models for the instrument noise. Similar ``glitch hypotheses'' were considered
by Veitch and Vecchio~\cite{Veitch:2009hd} in a study of Bayesian model selection applied to the search
for black hole inspiral signals. The possibility of subtracting instrument glitches from the data
using a physical model of the detector has been investigated by
Ajith {\it et al.}~\cite{Ajith:2007hg,Ajith:2014aea}. Principe and Pinto~\cite{Principe:2008bz} have
introduced a physically motivated model for the glitch contribution to the instrument noise, and have
used this model to derive what they refer to as a ``locally optimum network detection
statistic''~\cite{Principe:2009zz}. In their approach the glitches are treated in a statistical sense,
while our approach directly models the glitches present in the data. The possibility of
directly deriving likelihood functions from the data in a Bayesian setting has previously been
considered by Cannon~\cite{Cannon:2008zz}. In Cannon's approach the data is first reduced to an n-tuple
of quantities such as the parameters produced by a matched filter search.  Then, using time slides and
signal injections, the likelihood distributions for the signal and noise hypotheses are directly estimated
from the data. These are then used to estimate the posterior probability that a measured set of
parameters corresponds to a gravitational wave event.

The {\em BayesWave} algorithm models the data ${\bf s}$ as linear combination of gravitational wave signals, ${\bf h}$;
stationary, colored Gaussian noise, ${\bf n}_G$; and glitches, ${\bf g}$:
\begin{equation}\label{bwmodel}
{\bf s} = {\bf R}\star {\bf h} + {\bf g} + {\bf n}_G \, .
\end{equation}
The spectrum of the Gaussian component ${\bf n}_G$ is modeled by the {\em BayesLine} algorithm, running on
relatively short stretches of data (tens to hundreds of seconds) to allow for broad spectrum, long term non-stationary behavior.
We have experimented with using more complicated noise models for ${\bf n}$, such as the sum of Gaussians proposed by Allen {\it et al.}~\cite{Allen:2002jw,Allen:2001ay},
but found little benefit from the added complexity~\cite{Littenberg:2010gf}. 
Studies of the LIGO/Virgo data show that glitches are structured features that occupy
a compact region in the time-frequency plane. It is natural therefore to use some kind of wavelet representation for
the glitches. This could be a continuous wavelet representation such as the Morlet-Gabor ``Sine-Gaussian'' wavelets used in the LIGO Omega pipeline~\cite{omega05}
and for performing time-frequency decomposition of the data via Q-scans~\cite{omega05}, or a discrete wavelet representation such as those used in the Kleine-welle studies of
environmental channels or the coherent WaveBurst burst search pipeline~\cite{Klimenko:2004qh, cla, cWB}. Principe and Pinto~\cite{Principe:2008bz} have
shown that typical glitches can be represented as the sum of a small number of  Morlet-Gabor ``glitch atoms''. The {\em BayesWave} algorithm uses a
trans-dimensional RJMCMC to model glitches and signals with a variable number of Morlet-Gabor wavelets, where the number and parameters of the wavelets
are not pre-specified, but rather, determined from the data.

In principle the RJMCMC algorithm can produce posterior distributions for the components of the nested model (\ref{bwmodel}) which can be used for model selection.
For example, we can define a Gaussian noise model to be the state that employs zero glitch or signal wavelets. Similarly, we can define a pure glitch model to be one
that uses zero signal wavelets, and likewise, a pure signal model to be any state with zero glitch wavelets. In practice this is only useful for cases where the relative evidence
for each model (the pairwise Bayes Factors) are within a factor of a thousand or so. Beyond that point, the less favored models are visited so infrequently that the RJMCMC derived
Bayes Factors become suspect. It is possible to extend the RJMCMC analysis to higher Bayes Factors by imposing a false prior that weights the less favored model more heavily,
then correct for the false prior when computing the Bayes Factor~\cite{RJp}, but it gets increasingly difficult to achieve good mixing. To handle situations with large Bayes Factors we use thermodynamic integration to directly calculate the evidence for each model. The various models - Gaussian noise, glitches plus Gaussian noise and gravitational wave signal plus Gaussian noise - are themselves composite models formed by the union of all possible spectral estimates from {\em BayesLine}, and all possible combinations of wavelets in the glitch or signal models. Each composite model is explored by a RJMCMC, and thermodynamic integration is used to compute the evidence for the model. We are unaware of any other applications where RJMCMC and
thermodynamic integration have been combined in such a fashion. The evidence is used to select the model that best describes the data.

\section{Glitch and Signal Modeling}

Earlier versions of the {\em BayesWave} algorithm~\cite{Littenberg:2010gf, pbaker} used orthogonal bases of discrete Meyer wavelets~\cite{Klimenko:2004qh}. Initially we used a dyadic
basis uniformly spaced in log frequency~\cite{Littenberg:2010gf}, and later we experimented with a binary basis uniformly spaced in the frequency~\cite{pbaker}. The parameters
in these models were the number, amplitude and time-frequency location of the ``active'' wavelets. There are several advantages to using a discrete wavelet grid: the
parameters of the wavelet model are completely uncorrelated, resulting in good mixing of the Markov Chains; and it is easy to identify clusters of wavelets. There
are also several disadvantages: the shape of the pixels in time-frequency are fixed in advance by the decomposition level of the wavelet transform, leading to sub-optimal
representation of signals/glitches~\cite{pbaker}; and projecting the signal wavelets onto each detector in the network required the application of a computationally expensive time translation
operator~\cite{pbaker}.  We eventually concluded that the negatives outweighed the positives and abandoned discrete wavelets in favor of non-orthogonal
Morlet-Gabor wavelets which have a number of nice properties: their shape in time-frequency is variable and can adapt to fit the signal; they have an analytic Fourier representation so
the entire analysis can be performed in the frequency domain; and the time shifts needed to project the signal wavelets onto the various detectors are described by a simple phase
shift. The lack of orthogonality is not a major impediment to achieving good mixing in the Markov Chains, and clusters can be defined using a metric distance on the wavelet space.

Morlet-Gabor wavelets can be expressed in the time-domain as
\begin{equation}
\Psi(t; A, f_0, Q, t_0, \phi_0) = A e^{-(t-t_0)^2/\tau^2} \cos(2 \pi f_0(t-t_0) + \phi_0) \, ,
\end{equation}
with $\tau = Q/(2\pi f_0)$. Here $A$ is the amplitude, $Q$ is the quality factor, $t_0$ and $f_0$ are the central time and frequency, and $\phi_0$ is the phase offset.
Their Fourier transform is given by
\begin{eqnarray}
 \Psi(f; A, f_0, Q, t_0, \phi_0) &=& \frac{\sqrt{\pi} A\tau }{2} e^{-\pi^2 \tau^2 (f-f_0)^2} \times \nonumber \\
&&  \left(e^{i(\phi_0+2\pi(f-f_0)t_0)} + e^{-i(\phi_0+2\pi(f+f_0)t_0)} e^{-Q^2 f /f_0}\right).
\end{eqnarray}
Since we typically restrict $Q > 1$, the wavelets are highly peaked around $f=f_0$, and we may neglect the second term in the Fourier transform.
Morlet-Gabor wavelets have the minimum time-frequency area, $\pi \sigma_t \sigma_f = 1/2$, allowed by the Heisenberg-Gabor uncertainty
principle (their form is closely related to coherent states in quantum mechanics~\cite{grossman}). The glitch model employs a variable number of independent wavelets, $N_{g,i}$ in
each detector. The signal model can either use two sets of independent wavelets to model $h_+$ and $h_\times$ at the Geocenter, or for signals with a definite polarization, a single
set of wavelets to model $h_+$, and an ellipticity parameter $\epsilon$.  In the frequency domain we define $h_\times = \epsilon\, h_+ e^{i\pi/2}$,
with the polarization going from linear to circular as  $\epsilon$ runs from 0 to 1.  Our motivation for assuming elliptical polarization is two-fold: Firstly, the
the small number of detectors available in the early advanced detector era, and the near alignment of the two LIGO detectors, results in poor polarization sensitivity,
making it difficult to reconstruct $h_+$ and $h_\times$ separately. Secondly, we expect most sources of gravitational waves to show some degree of polarization.
For example, the signals from binary inspirals are elliptically polarized with ellipticities that are approximately constant across multiple wave cycles, while
signals from core collapse supernovae are thought to be dominantly linearly polarized. In what follows we will assume that the signals are polarized, and that $\epsilon$ is time
independent. The Geocenter signal wavelets are projected onto each detector in the network using sky-location-dependent time-delays $\Delta t(\theta,\phi)$,
and antenna patterns $F^+(\theta, \phi, \psi)$, $F^\times(\theta, \phi, \psi)$:
\begin{equation}
({\bf R}\star {\bf h})_i(f) = \left(F_i^+(\theta, \phi, \psi) h_+(f) + F_i^\times(\theta, \phi, \psi) h_\times(f)\right) e^{2\pi i f \Delta t_i(\theta,\phi)} \, ,
\end{equation}
where
\begin{equation}
h_+(f) = \sum_{j=0}^{N_{s}} \Psi(f; A_j, f_{0j}, Q_j, t_{0j}, \phi_{0j}) \, .
\end{equation}
All $N_s$ wavelets in the signal model share the same set of extrinsic parameters describing the sky location $(\theta,\phi)$, ellipticity $\epsilon$, and polarization angle $\psi$.
A gravitational wave signal can be fit equally well (in terms of the likelihood) by the glitch or signal models, but the evidence for the signal model will be larger as it provides a
more parsimonious description of the data. For a signal described by $N_s$ wavelets incident on a $N_d$ detector network, the signal model uses $D_s=5N_s + 4$ parameters
(the 5 extrinsic parameters per wavelet and the 4 common extrinsic parameters), while the glitch model uses $D_g = 5 N_d N_s$ parameters. When $N_s \gg 1$ or $N_d \gg 1$,
the glitch model requires vastly more parameters than the signal model. When $N_s = 1$ and $N_d = 1$ the extrinsic parameters are irrelevant and $D_s=D_g$, which makes
it impossible to tell the difference between an un-modeled gravitational wave signal and a glitch using a single detector. The most marginal case for detection is a two
detector network and a single sine-Gaussian signal, as then $D_s=9$ and $D_g=10$. However, despite the near equality of the model dimensions in this case, the signal model will still
be preferred as not all parameters are created equal - models pay a greater penalty for parameters that are well constrained (measured by the ratio of prior to posterior volume). Intrinsic parameters such as $t_0$ and $f_0$ fall into this category.

\section{Priors}

Having defined the likelihood and the building blocks of the signal and glitch models, all that remains is to specify prior distributions on the model parameters.
For the extrinsic parameters of the signal model we adopt the prior that sources are uniformly distributed on the sky, with ellipticities uniform in the range $\epsilon \in [0,1]$,
and polarization angles uniform in the range $\psi \in [0,\pi]$. Similarly, we take the prior on the intrinsic parameters $t_0$ and $f_0$ to be uniformly distributed
across the observation time and frequency band being analyzed. These priors should be modified if we have an electromagnetic counterpart such as a core collapse
supernovae, which would pin down the sky location and perhaps the ellipticity, and narrow the range on $t_0$. The phase parameters are taken to be uniform in the range
$\phi_0 \in [0,2\pi]$. We take the quality factors to be uniform in the range $Q \in [Q_{\rm min}, Q_{\rm max}]$, where we typically set $Q_{\rm min} =2 $ and $Q_{\rm max} = 40$.
We adopt a uniform prior on the number of wavelets in the glitch and signal models across the range $N_s \in [0,100]$ and
$N_g \in [0, 100\times N_d]$.  The prior range on the quality factor and number of wavelets was informed by running the analysis on a variety of simulated signals, such as black hole binary mergers,
core collapse supernovae and white noise bursts, and on glitches in actual LIGO data and ensuring that the parameters used by the models lay well within the boundaries of the prior range.
When running on the signal model the ranges are set such that $N_s \in [1,100]$, $N_g=0$, while the glitch model has $N_s = 0$ and $N_g\in [1,100\times N_d]$ (in other words, a single glitch wavelet
in any detector is acceptable). The Gaussian noise model has $N_g=0$ and $N_s=0$. All that remains is to specify an amplitude prior for the wavelets. In this instance, a uniform prior has little to recommend it,
and we need to specify physically well motivated priors for the signal and glitch amplitudes. In each case we specify a prior on the amplitude of individual wavelets, but these may be supplemented by priors on the
total power in the signal and glitch models.

\subsection{Glitch amplitude prior}
From previous studies we know that loud glitches are less common than quiet glitches. Once the amplitude of a glitch becomes comparable to the Gaussian noise level it loses
identity and becomes one of the many small disturbances that make up the Gaussian background. These considerations motivate an amplitude prior expressed in terms of the signal-to-noise ratio (SNR) of the wavelet. For glitches we adopt the form
\begin{equation}\label{aprior}
p({\rm SNR}) = \frac{{\rm SNR}}{{\rm SNR}^2_*} \; e^{-{\rm SNR}/{\rm SNR_*}} 
\end{equation}
which peaks at ${\rm SNR}={\rm SNR}_*$. The ${\rm SNR}$ of a Morlet-Gabor wavelet can be estimated as
\begin{equation}\label{SNRe}
{\rm SNR} \simeq \frac{A \sqrt{Q}}{\sqrt{2 \sqrt{2 \pi} f_0 S_n(f_0)}} \, ,
\end{equation}
which means that the ${\rm SNR}$ prior is actually a joint prior on $A, f_0, Q$. A non-flat prior on the ${\rm SNR}$ will result in a bias in the recovered amplitude
which is tolerable so long as it is not too large. To estimate the bias, consider a single sine-Gaussian signal with signal-to-noise $\overline{{\rm SNR}} = (s\vert s)^{1/2}$, and further
assume that the template exactly matches the signal in all parameters other than the amplitude. The posterior distribution for ${\rm SNR}$ then has the form
\begin{equation}
p({\rm SNR}\vert s) \sim  {\rm SNR} \; e^{-{\rm SNR}/{\rm SNR_*}}  \; e^{-(\overline{{\rm SNR}} - {\rm SNR})^2/2}  \,  ,
\end{equation}
Note that for large signal-to-noise, the likelihood overwhelms the prior, and in the limit $\overline{{\rm SNR}} \gg  1$ the posterior distribution peaks at
\begin{equation}
{\rm SNR} \simeq \overline{{\rm SNR}} \left( 1 - \frac{1}{\overline{{\rm SNR}}\, {\rm SNR}_*}\right) \, .
\end{equation}

\begin{figure}[ht]\label{fig:bias}
\centerline{\includegraphics[scale=1.0]{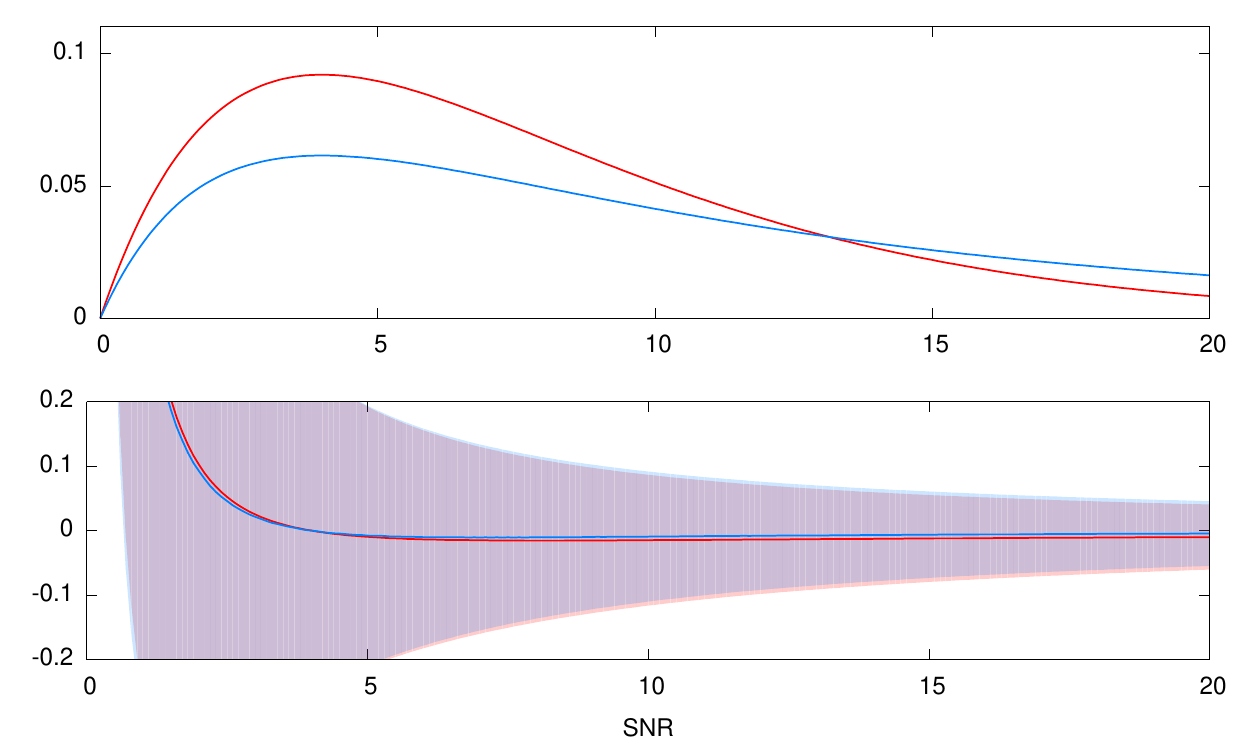}}
\caption{The upper panel shows the SNR prior for the glitch wavelets (in red), defined in Eq. (\ref{aprior}), and the signal wavelets (in blue), defined in Eq. (\ref{asprior}), for the choice
${\rm SNR_*}=4$. The lower panel shows the fractional
bias in the inferred SNR caused by using these non-uniform SNR priors. The shaded regions fold in the 1-sigma statistical error in the SNR estimates, which are always far greater
than the systematic biases from the non-uniform priors.}
\end{figure}

\subsection{Signal amplitude prior}
For gravitational wave signals we expect the sources to be distributed roughly uniformly in volume, which implies a prior on the distance $D$ that scales as $p(D) \sim D^2$.
Since the SNR of a signal scales as ${\rm SNR} \sim 1/D$, the corresponding prior on SNR scales as $p({\rm SNR}) \sim {\rm SNR}^{-4}$. One possibility is to use this as
the prior on the individual wavelet SNRs, but such a prior is improper (can not be normalized), and it favors adding large numbers of undetectable wavelets to the signal model.
While we have reservations about employing so-called ``Malmquist priors''\footnote{We feel that Malmquist priors imply a double counting of what is already contained in
the likelihood, though they may be necessary if the parameter estimation study is following up triggers from a search pipeline with known selection effects~\cite{Rover:2007ij}},
that seek to account for observational selection effects, convergence can be greatly improved by cutting off the distribution at some minimum SNR. This does not affect the
ability to reconstruct signals since the natural balance between model fidelity and model complexity ensures that only wavelets with ${\rm SNR} > 3$ play a
significant role. Rather than adopt a hard SNR cut-off, we instead choose a function that smoothly
goes to zero as ${\rm SNR} \rightarrow 0$, and approaches the ${\rm SNR}^{-4}$ scaling for high SNR. One such choice is
\begin{equation}\label{asprior}
p({\rm SNR}) =    \frac{ 3\, {\rm SNR}}{ 4 \, {\rm SNR}_*^2 \left(1+{\rm SNR}/(4 \, {\rm SNR}_*)\right)^5} \, .
\end{equation}
This distribution peaks at ${\rm SNR} = {\rm SNR}_*$. The SNR of a signal wavelet can be estimated using (\ref{SNRe}), but with the power spectral density of
a single detector $S_n(f_0)$ replaced by the network average
\begin{equation}
\bar{S}_n(f_0)  = \left( \sum_i  \frac{F_{+,i}^2 + \epsilon^2 F_{\times,i}^2}{S_{n,i}(f_0)} \right)^{-1} \, .
\end{equation}
Thus our ${\rm SNR}$ prior for the signal model is actually a joint prior on $A, f_0, Q, \epsilon, \psi, \theta,\phi$. As with the glitch model, there will be a bias away from the maximum likelihood
solution for the inferred SNR caused by the non-uniform prior on SNR. Figure~1 compares the glitch and signal SNR priors for the choice ${\rm SNR}_*=4$, and shows the fractional bias in
the inferred SNR caused by using a non-uniform prior. In both cases, the bias is tiny compared to the statistical uncertainty.

\subsection{Cluster Priors}

In addition to the individual wavelet priors described above, it is possible to introduce non-local priors that apply to collections of wavelets. On physical grounds we
expect gravitational wave signals, such as black hole mergers or core collapse supernovae, and glitches to produce coherent clusters of power in a time-frequency representation.
This suggests imposing priors that favor concentrated
clusters of wavelets over isolated wavelets.

\begin{figure}[ht]\label{fig:cnum}
\centerline{\includegraphics[scale=1.0]{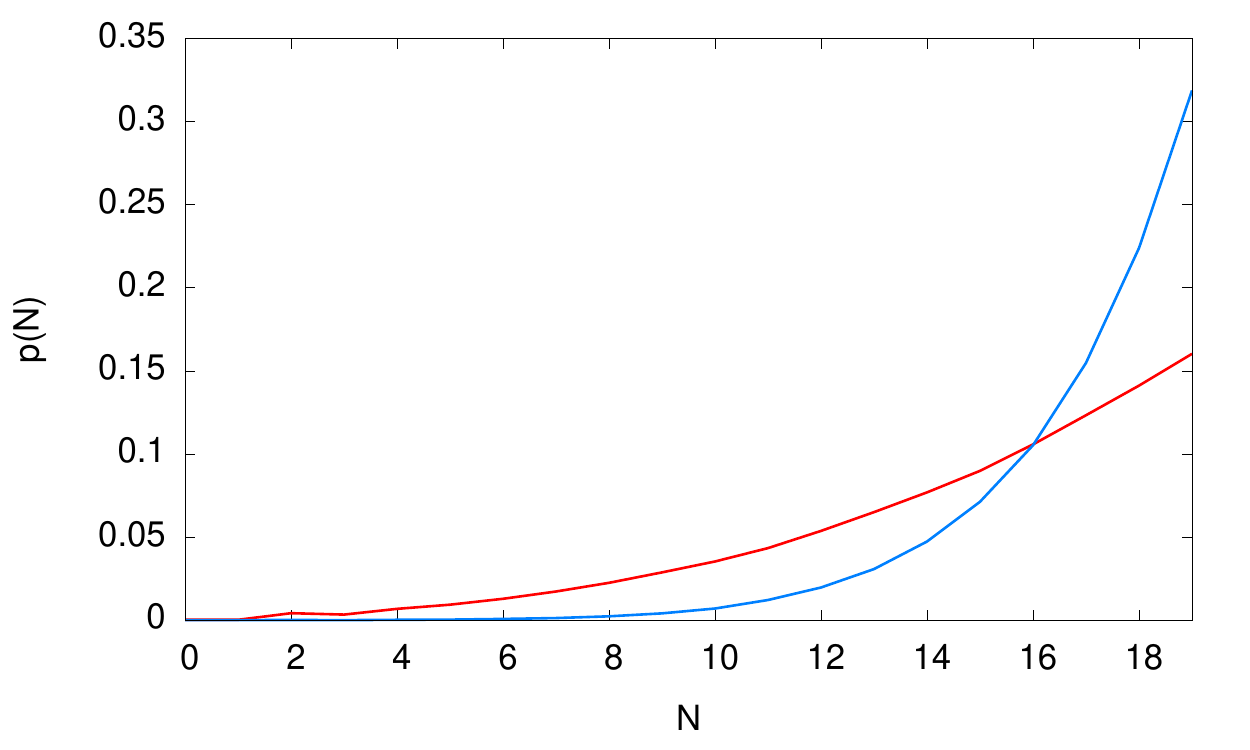}}
\caption{Distributions of the number of active wavelets when using the proximity prior defined in Eq. (\ref{proxfull}) with the likelihood set equal to a constant.
The blue line is for $(\alpha,\beta,\gamma)=(4,1,0.5)$ and the red line
is for $(\alpha,\beta,\gamma)=(2,0.5,0.5)$. To achieve a uniform prior distribution on the number of wavelets, we include a normalization factor proportional to the inverse of these distributions.}
\end{figure}

One approach that we considered uses a distance measure to decide which wavelets are nearby in the time-frequency
plane, then groups neighboring wavelets into a cluster. Treating the wavelets as Gaussian packets in time-frequency space with density
\begin{equation}\label{den}
\rho(t,f) = \frac{1}{2 \pi\sigma_f \sigma_t}e^{-(\Delta f^2/2\sigma_f^2+\Delta t^2/2\sigma_t^2)},
\end{equation}
where  $\Delta f = f_0-f$, $\Delta t = t_0-t$, $\sigma_t = \tau$ and $\sigma_f = 1/\pi \tau$, and computing the overlap between two wavelets yields the
distance measure
\begin{equation}
ds^2_{12} = \frac{ dt^2 + (\pi \tau_1 \tau_2)^2 df^2}{\tau_1^2 + \tau_2^2} \, ,
\end{equation}
where $dt = t_{0_1}-t_{0_2}$ and $df= f_{0_1}-f_{0_2}$ is the difference in the central time and central frequencies between wavelets $1,2$.
We then set a distance threshold such that all wavelets within $ds < \delta$ of each other were considered to be in a cluster, and used the {\em Flood Fill} algorithm~\cite{floodfill} to identify clusters.
From this we were able to extract the number of clusters, the number of wavelets in each cluster, and the signal-to-noise ratio of each cluster.  We
experimented with a number of priors that used these quantities to favor groupings of wavelets, but each attempt ran into various difficulties. In general there was a strong tendency to
form large clusters, even in Gaussian noise, and these clusters would typically have wavelets with extreme $Q$'s producing an unphysical mix of both very tall and very long pixels
that linked the clusters together. The tendency to over-cluster could be balanced to a certain extent by adding a prior that disfavored configurations with a large number of wavelets,
but the bias towards extreme wavelet shapes remained.

\begin{figure}[ht]\label{fig:prior}
\centerline{\includegraphics[scale=1.0]{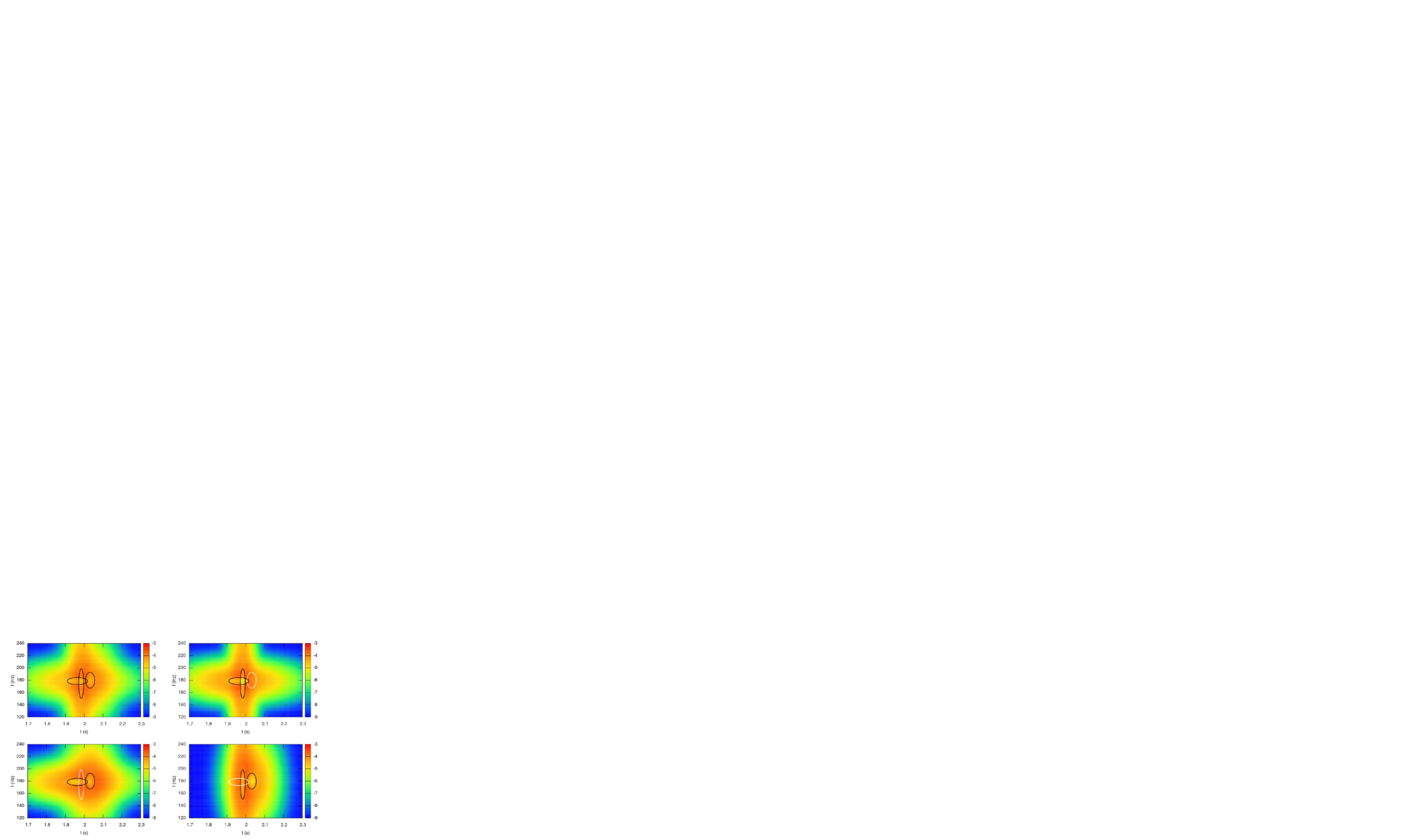}}
\caption{Density plots in time-frequency showing the log of the proximity prior with $(\alpha,\beta,\gamma)=(4,1,0.5)$ taken from a MCMC run on data containing a Gaussian enveloped white noise
burst with ${\rm SNR}=15$, central frequency $f=180$ Hz, central time $t = 2$ s and spreads $\sigma_f=20$ Hz, $\sigma_t=0.1$ s. The Maximum a Posterior wavelet model employed 3 wavelets. The upper left
panel shows the log of the proximity prior density overlaid with the shapes of the three wavelets in use at a randomly chosen iteration of the Markov chain. The other three panels show the prior as
seen by the wavelet colored in white, which are found by excluding the contribution to the prior from that wavelet.}
\end{figure}

The approach we settled on employs a ``proximity prior'' based on a sum of probability densities surrounding each active wavelet. Since each wavelet covers a region of the time-frequency
plane defined by the density (\ref{den}), we expect neighboring wavelets to be close, but not too close. This motivates a probability density described by a ``hollowed-out'' Gaussian of the form
\begin{equation}\label{prox}
\fl \quad \quad p(t,f) = \frac{1}{2 \pi\sigma_f \sigma_t(\alpha^2-\beta^2)}\left(e^{-(\Delta f^2/\sigma_f^2+\Delta t^2/\sigma_t^2)/\alpha^2}-e^{-(\Delta f^2/\sigma_f^2+\Delta t^2/\sigma_t^2)/\beta^2}\right) ,
\end{equation}
where $\alpha > \beta$ sets the overall width of the distribution and $\beta$ sets the size of the hollowed-out region. We expect that choosing $\alpha$ in the range 2-4 and $\beta$ in the range 0.5-1
will result in an effective proximity prior for most signal and glitch morphologies.  An extensive study of how these choices affect the performance is currently underway. The full proximity prior
is given by the normalized sum of densities of the form  Eq. (\ref{prox}) surrounding each active wavelet, to which we also add a uniform density component to allow exploration of the entire time-frequency
volume:
\begin{equation}\label{proxfull}
\fl \quad \quad p(t,f) = \sum_{i =1}^{N} \frac{(1-\gamma)}{N A_i} \left(e^{-(\Delta f_i^2/\sigma_f^2+\Delta t_i^2/\sigma_t^2)/\alpha^2}-e^{-(\Delta f_i^2/\sigma_f^2+\Delta t_i^2/\sigma_t^2)/\beta^2}\right) +  \frac{\gamma}{V_{tf}} \, 
\end{equation}
where $V_{tf}$ is the time-frequency volume being analyzed, $\Delta f_i = f_{0_i}-f$, $\Delta t_i = t_{0_i}-t$, and $A_i$ is the normalization factor for the $i^{\rm th}$ hollowed-out Gaussian
distribution:
\begin{equation}\label{prox}
A_i =\int dt \ df \left(e^{-(\Delta f^2/\sigma_f^2+\Delta t^2/\sigma_t^2)/\alpha^2}-e^{-(\Delta f^2/\sigma_f^2+\Delta t^2/\sigma_t^2)/\beta^2}\right) \, .
\end{equation}
We set the uniform fraction at $\gamma$, and the hollowed-out Gaussian fraction is $1-\gamma$. Rather than choosing a fixed fraction $\gamma$, we instead choose a fixed density contrast
between the two components, which results in $\gamma$ taking different values depending on the time-frequency volume being analyzed. The proximity prior
favors states with large numbers of nearby wavelets. This can be seen most clearly in prior-only studies where the likelihood is set equal to a constant. We find that the proximity prior leads to
an approximately power law distribution in the number of active wavelets, as opposed to the uniform distribution in the number of active wavelets found when using a uniform time-frequency prior.
The precise form of the wavelet number distribution depends on the choice of $\alpha,\beta,\gamma$ and the size of the time-frequency volume being analyzed. Figure 2 shows the distribution
of the number of active wavelets in prior-only studies with $(\alpha,\beta)=(4,1)$ and $(\alpha,\beta)=(2,0.5)$. In each case the range in the number of active wavelets was set at $N=[0,19]$, and the analysis
covered the frequency range $f=[10,512]\, {\rm Hz}$ and a time interval of 8 seconds, resulting in a time-frequency volume of ${\rm TFV} = 4016$. The density contrast in the proximity prior was
set such that $\gamma = 1/(4016/{\rm TFV}+1)=0.5$. The more aggressive proximity prior with $(\alpha,\beta)=(2,0.5)$ leads to a steeper trend in the number of active wavelets being used. Since our
preferred physical prior on the number of active wavelets is uniform, we include a normalization factor on the number of wavelets that compensates for the over-clustering caused by the proximity prior.
The obvious choice is to use the inverse of the distribution shown in Figure 2, which results in a uniform distribution in the number of active wavelets when the likelihood is set equal to a constant.

Figure 3 shows the log of the proximity prior density for a cluster of wavelets randomly chosen from a MCMC analysis. The upper left panel shows what the prior would look like when
proposing to add an additional wavelet to the cluster. The other panels show what the prior looks like for each individual wavelet in the cluster {\it i.e.} what the prior looks like when
proposing to add the third wavelet (shown in white) to the cluster. Note that the dynamic range in the proximity prior between the peak value and the minimum is of order $\Delta \log p \approx 6$,
which means that the improvement in the log likelihood needed to add a new wavelet near the cluster is significantly lower than elsewhere in the time-frequency plane.

\begin{figure}[ht]\label{fig:prior_clust}
\begin{center}
    \includegraphics[scale = 0.8]{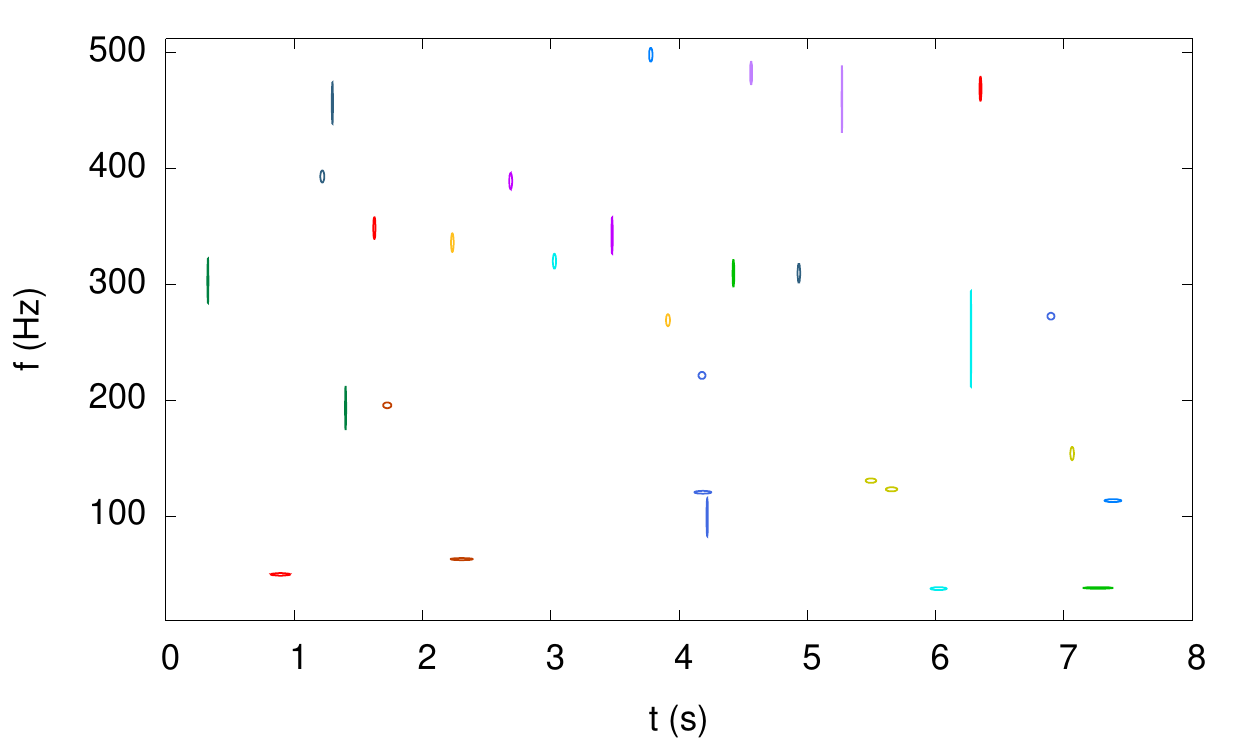}\\
    \includegraphics[scale = 0.8]{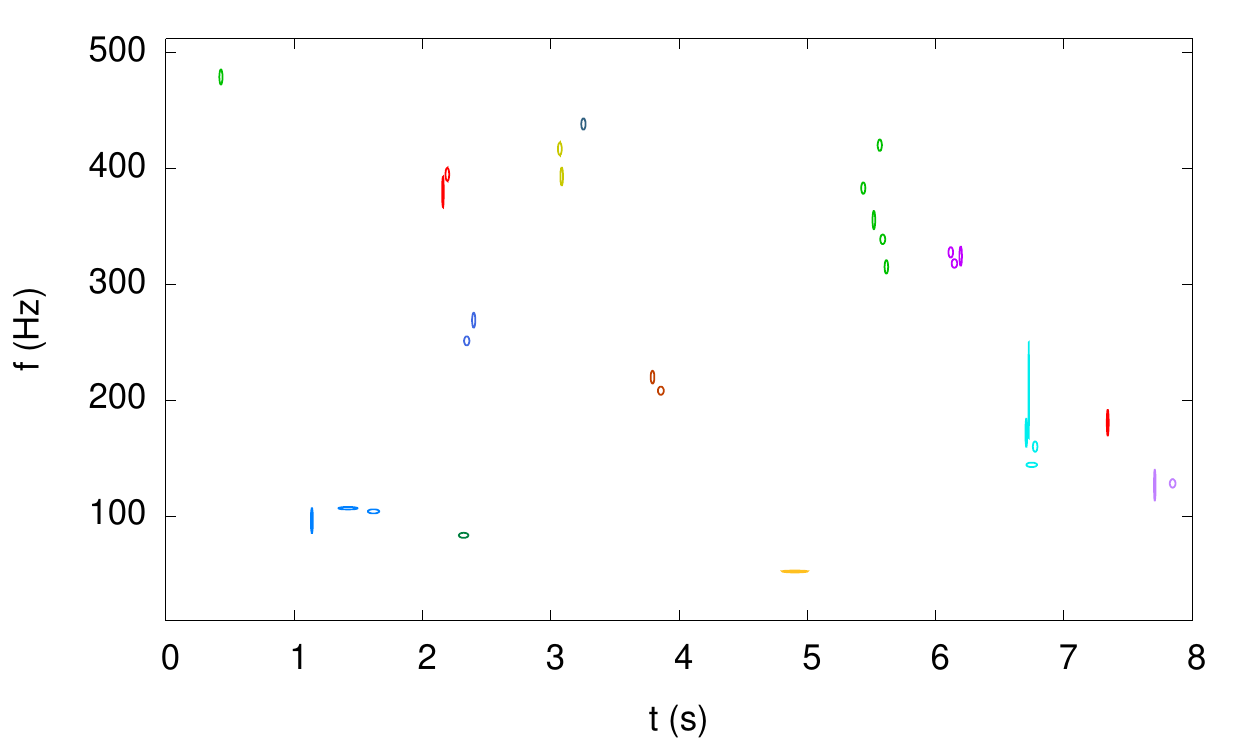}
  \end{center}
\caption{Wavelet clusters, indicated by a common color, in the time-frequency plane when exploring the prior distributions (likelihood set equal to a constant). These snapshots were randomly selected 30-wavelet
states taken from MCMC runs with a uniform time-frequency prior (upper panel) and the proximity prior with $\alpha=4,\beta =1,\gamma=0.5$ (lower panel). The
proximity prior leads to a larger number of multi-wavelet clusters.}
\end{figure}

Figure 4 shows randomly chosen snapshots of the the wavelet distributions for states with 30 wavelets from runs where the log likelihood was set equal to a constant. The wavelets are grouped into
clusters using the {\em Flood Fill} algorithm~\cite{floodfill} with a distance tolerance of $ds < 4$ and assigned a common color. Note that since there are only a finite number of colors in the color table being used to
make the plot, some wavelets get plotted in the same color even though they are not in the same cluster. The upper panel is with a uniform time-frequency prior, while the lower panel
is with the proximity prior. As expected, the proximity prior shows a distinct preference for forming clusters composed of many wavelets.

\begin{figure}[ht]\label{fig:match}
\centerline{\includegraphics[scale=1.0]{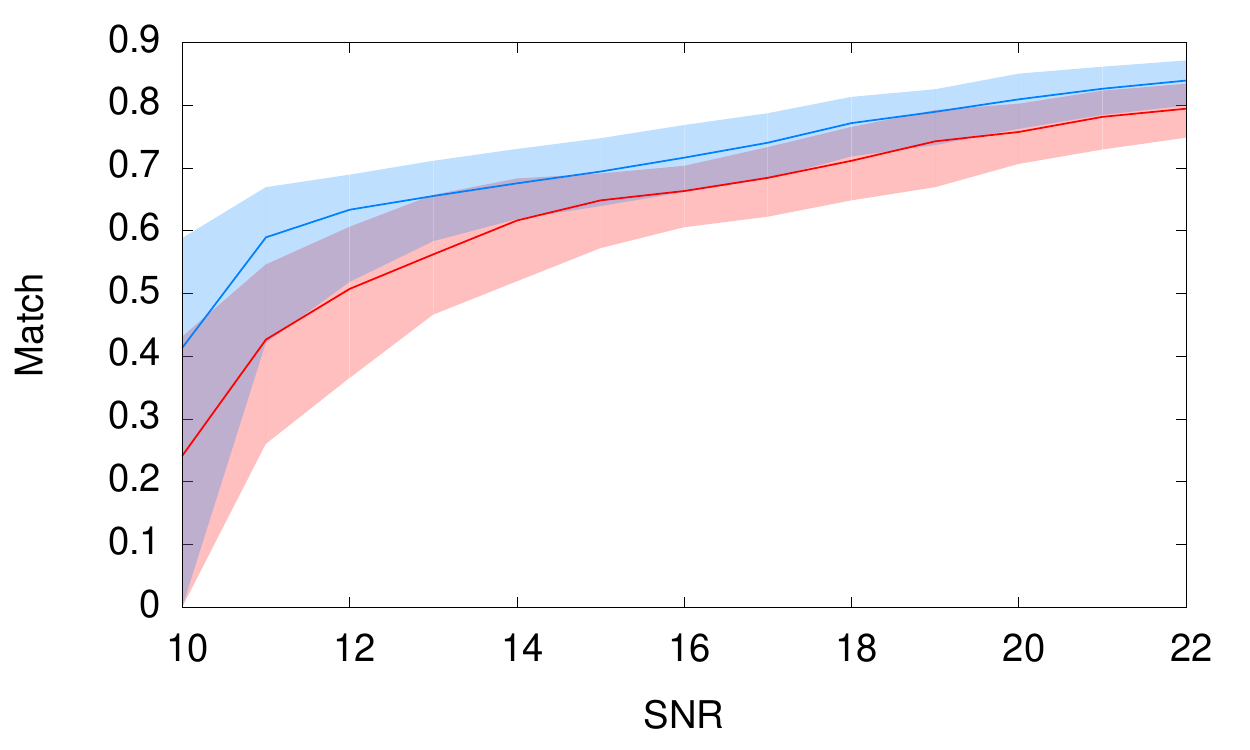}}
\caption{The median match for the wavelet model as a function of signal-to-noise ratio for a simulated black hole merger. The red line shows the median match and the symmetric 90\%
credible interval for a uniform time-frequency prior, while the blue line and shaded region shows the same quantities for the proximity prior with $\alpha=4,\beta =1,\gamma=0.5$. }
\end{figure}

Figure 5 shows quantiles of the match between an injected black hole merger signal $\bar{h}$ and the wavelet reconstruction $h$ found by {\em BayesWave} for a flat time-frequency prior and a proximity prior with
$\alpha=4,\beta =1,\gamma=0.5$.  The match, or overlap, is defined:
\begin{equation}\label{match}
{\rm M} = \frac{ (\bar{h} \vert h)}{\sqrt{(\bar{h} \vert \bar{h})(h\vert h)}} \, ,
\end{equation}
and can range between -1 and 1.
The black hole merger signal included inspiral, merger and ringdown for a system with spins aligned with the orbital angular momentum, and
spin magnitudes $\chi_1=0.3$, $\chi_2=0.2$ and masses $m_1 = 20\, M_\odot$, $m_2 = 15\, M_\odot$. The proximity prior with $\alpha=2,\beta =0.5,\gamma=0.5$ (not shown in the
interest of readability) produced matches that fell between the two cases shown in Figure 5. In the future we plan to perform a more extensive study of how the proximity prior
performs with different choices of $\alpha,\beta,\gamma$ for a variety of simulated signals. For the remainder of the paper we will adopt the fiducial values $\alpha=4,\beta =1,\gamma=0.5$.

\begin{figure}[ht]\label{fig:match}
\centerline{\includegraphics[scale=1.0]{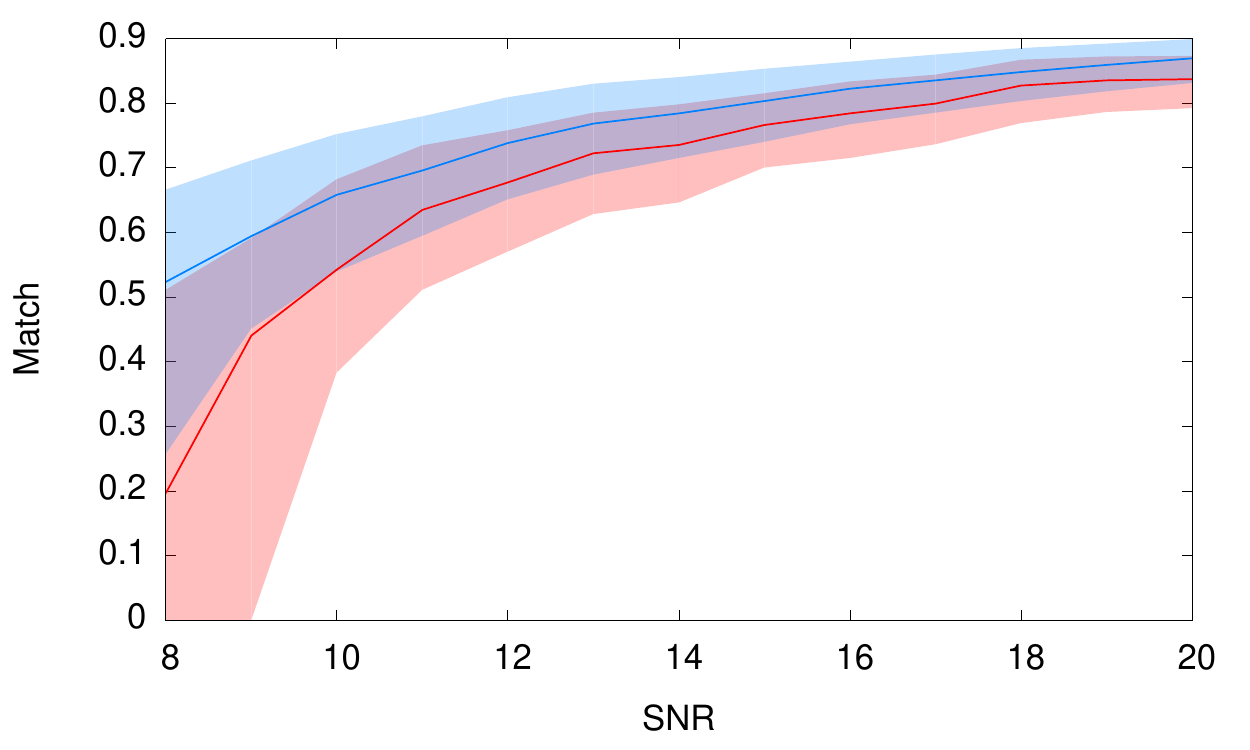}}
\caption{The median match for the wavelet model as a function of signal-to-noise ratio for a simulated white noise burst. The red line shows the median match and the symmetric 90\%
credible interval for a uniform time-frequency prior, while the blue line and shaded region shows the same quantities for the proximity prior with $\alpha=4,\beta =1,\gamma=0.5$}
\end{figure}

Figure 6 shows quantiles of the match as a function of signal-to-noise ratio for a simulated Gaussian enveloped white noise burst with central frequency $f=180$ Hz and spreads
$\sigma_f=20$ Hz, $\sigma_t=0.1$ s. Once again the proximity prior significantly outperforms the uniform time frequency prior in terms of the accuracy of the waveform recovery.

\section{RJMCMC Implementation}

The {\em BayesWave} algorithm uses a trans-dimensional RJMCMC technique~\cite{Green} to explore the models and their many parameters.
Transitions to new states of the chain are drawn from a proposal distribution, and to achieve efficient sampling the proposals need to be well adapted to the problem. The
most challenging transitions are those that involve adding or removing a wavelet (the trans-dimensional steps), and birth/death moves that simultaneously remove one wavelet and
add another. To ensure a high acceptance rate, {\em BayesWave} uses a variety of custom tailored proposal distributions and additional techniques to promote mixing.

\subsection{Maximization during Burn-in}
To speed convergence, standard FFT based techniques are used to maximize over $t_0$, $\phi_0$ and $A$ for each wavelet during the ``burn-in'' stage. Samples from this part of the chain are
discarded since they do not obey the detailed balance condition required for Markov chains. The maximization allows signals and glitches to be identified very quickly.

\subsection{Parallel Tempering}
Convergence to the target distribution is improved by using the Parallel Tempering approach~\cite{PT}, where multiple chains are run in parallel at different ``temperatures'' $T$, where the
likelihood function is modified such that $p({\bf s}\vert {\bf h}) \rightarrow p({\bf s}\vert {\bf h})^{1/T}$. Hot chains explore the entire prior volume, while cooler chains map the
peaks of the posterior distribution. Exchanges between chains at different temperatures help ensure that the entire posterior distribution is thoroughly explored. The choice of temperature
ladder can significantly impact the mixing between chains. We typically use 25-30 chains with the temperature spacing determined by the adaptive algorithm described in Ref.~\cite{Vousden} and the
maximum temperature fixed to $T=10^6$.

\subsection{Fisher matrix proposals}
A Fisher information matrix $\Gamma$ can be computed for each wavelet, and proposals for updating the parameters of that wavelet can be drawn from the multivariate Gaussian distribution
\begin{equation}
q({\bf y}\vert {\bf x})= \left(\frac{{\rm det}\Gamma}{(2\pi)^2}\right)^{1/2} e^{-\frac{1}{2} \Gamma_{ij}\Delta x^i \Delta x^j},
\end{equation}
where the $\Delta x^i = x^i - y^i$ denote displacements in the five intrinsic parameters that describe each wavelet. To speed calculation we use an analytic approximation to the Fisher matrix that
can be derived along the same lines as the SNR in (\ref{SNRe}). Dropping terms down by factors of $e^{-Q^2}$ relative to leading order, the Fisher matrix for a single wavelet using the
parameters $\{t_0, f_0, Q, \ln A,  \phi_0\}$ is given by
\begin{equation}
\Gamma = {\rm SNR}^2 \left( \begin{array}{ccccc}
\frac{4 \pi^2 f_0^2 (1+Q^2)}{Q^2} & 0 & 0 & 0  & -2\pi f_0\\
0 & \frac{3 +Q^2}{4f_0^2} & -\frac{3}{4 Q f_0} & -\frac{1}{2 f_0} & 0 \\
0 &  -\frac{3}{4 Q f_0}  &  \frac{3}{4 Q^2}  & \frac{1}{2 Q} & 0 \\
0  &  -\frac{1}{2 f_0} & \frac{1}{2 Q}  & 1 &  0 \\
-2\pi f_0 & 0  & 0 & 0 & 1
\end{array} \right)
\end{equation}
Jumps are proposed along the eigenvectors of this matrix, scaled by the inverse square root of the eigenvectors. The eigenvectors come in two groups, one group that mixes
$t_0$ and $\phi_0$, and another group that mixes $f_0,Q$ and $\ln A$. While these jump proposals do not account for correlations between the parameters of different wavelets, the
correlations between wavelets are small, and the single wavelet Fisher matrix based jumps achieve a healthy acceptance rate of $\sim 30\%$.

\subsection{Extrinsic parameter proposals}
Changes to the extrinsic parameters which describe the location and orientation of the source are proposed using a mixture of three distributions.  The most heavily employed proposal uses the Fisher information matrix of the Geocenter waveform similar to what is used for the intrinsic parameters.  There is no simple analytic expression for derivatives of the signal model with respect to extrinsic parameters, so the elements of the Fisher matrix are computed by finite differencing.  Numerical differentiation is computationally costly so the Fisher matrix is treated as constant over several chain iterations (typically ten) before being updated.

The Fisher matrix proposal assumes a multivariate Gaussian likelihood which is a poor approximation for parameters that encode the position and orientation of the signal.  Ground-based gravitational wave detector networks acquire most of the positional information by differences in the arrival time of the signal at each observatory.  For two detector networks (the minimum required for confident detection) timing-only considerations produce a degenerate ring on the sky of constant time delay.  This degeneracy is broken by the different phase and amplitude measured in each detector but the resulting distributions are not remotely Gaussian (for example, see
Figure~15).  We incorporate this known feature of gravitational wave posteriors into a proposal which rotates the current sky-location uniformly along the ring of constant time delay.  

Rounding out the extrinsic parameter proposals is a uniform draw from the prior on all parameters.  Proposals which make such large changes to the extrinsic parameters are seldom, if ever, accepted by the cold chains but are necessary for the hotter chains in the parallel tempering scheme to ensure that they fully sample the prior.  The full prior exploration is important to help the cold chains avoid getting trapped in local maxima of the posterior and to get an accurate estimate of the expectation value of the likelihood as a function of temperature that is used in the thermodynamic integration.

\subsection{Trans-dimensional and birth-death moves}
When proposing to add a new wavelet, or swapping out an existing wavelet, the choice of time-frequency location for the proposed wavelet is key to getting the move accepted. {\em BayesWave}
employs two kinds of time-frequency proposals to help facilitate these moves. The first is based on a time-frequency map which favors adding wavelets in regions with excess power.
The second is a proximity proposal, which preferentially seeks to add new wavelets near to, but not on top of, existing wavelets.

The time-frequency map can be constructed in a variety of ways. One way is to
use a normalized discrete scalogram of the whitened data in each detector. The glitch model in each detector uses the whitened scalogram for that detector, while the signal model uses
the average of the scalograms for each detector in the network. Another way of constructing a time-frequency map is to perform a pilot run for each model (signal and glitch) and form a histogram of where
the wavelets get placed in time-frequency space. Both methods for constructing time-frequency maps have proven to be successful in testing. The current implementation of the code uses the
histogram based approach. 

The proximity proposal distribution takes the same functional form as the proximity prior, though possibly with different values for the parameters $(\alpha,\beta,\gamma)$. The idea behind this proposal is
that new wavelets are more likely to be needed in the vicinity of an existing cluster. In the studies presented here the proximity proposal used the same settings as the proximity prior, with
 $\alpha=2,\beta =0.5,\gamma=0.5$.

Once the central time $t_0$ and frequency $f_0$ have been drawn from either the time-frequency map or proximity proposal, values for the other wavelet parameters need to be proposed.
The quality factor $Q$ and phase $\phi_0$ are drawn from their uniform prior range, while the amplitude $A$ of the new wavelet is proposed by drawing a value for the proposed ${\rm SNR}$ from
the amplitude prior (\ref{aprior}), and solving for $A$ by inverting the expression in (\ref{SNRe}). For wavelets in the signal model, the extrinsic parameters (sky location, polarization angle and ellipticity)
are inherited from the existing signal model wavelets, or if the signal model currently has no wavelets, these are drawn from their prior distributions. The Metropolis-Hastings ratio for the
trans-dimensional moves includes many non-trivial terms in the prior and proposal densities, and care has to taken when implementing the algorithm. The implementation is
tested by setting the likelihood equal to a constant and seeing if the algorithm recovers the prior distributions. It is hard to over-stress how useful and important this test is when developing a
code as complex as {\em BayesWave}.

\section{Bayesian Evidence}

In order to compare various models for the data we need to be able to compute the posterior probability for each model.
The (un-normalized) posterior probability for a particular model $M$ is given by the product of its prior odds $p(M)$ and its marginal likelihood or evidence $p(s\vert M)$. The marginal likelihood
for a model is given by an integral of the model likelihood weighted by the prior distribution of the model parameters:
\begin{equation}
p(s\vert M) = \int p(\vec{\lambda}\vert M) p(s\vert \vec{\lambda}, M) d \vec{\lambda} \, .
\end{equation}
This integral is notoriously difficult to evaluate when the model has a large number of parameters. Several ingenious techniques have been developed to estimate the evidence,
including nested sampling and thermodynamic integration. An alternative approach is to avoid a direct calculation of the evidence and instead use a trans-dimensional RJMCMC algorithm to
simultaneously explore both the space of models and the parameters of each model. The probability for each model can then be estimated from the relative frequency of visits to each model.

{\em BayesWave} uses a RJMCMC routine to explore the full range of models (gravitational wave signals, glitches and Gaussian-noise), so it would seem that we get the model probabilities for
free. We can define four disjoint composite models: (1) Gaussian-noise only; (2) Gaussian-noise and glitches - states with one or more glitch wavelets in use across the network; (3) Gaussian-noise and
a gravitational wave signal - with one or more signal wavelets in use; (4) Gaussian-noise, glitches and a gravitational wave signal - with one or more signal wavelets and one or more glitch wavelets in use; and use
the fraction of iterations spent in each model as an estimate of the posterior probability for each model. The challenge is to get the RJMCMC algorithm to efficiently explore each composite model
and to transition between models. It is worth emphasizing that the composite models are of variable dimensionality - the Gaussian noise model allows for a variable number of spline control points
and Lorentzian line features, and the glitch and gravitational wave models allow for a variable number of wavelets. Thus the RJMCMC algorithm has to marginalize over the internal model degrees of freedom
in addition to exploring the composite model space. In situations where there are near-coincident loud glitches in multiple detectors or a loud gravitational wave signal, it becomes very difficult to come up with
proposals that allow for efficient transitions between the all-glitch and all-signal models. One way around this is to do pairwise comparisons between a sub-set of models, such as Gaussian-noise versus
Gaussian-noise and glitches or Gaussian-noise versus Gaussian-noise and gravitational waves. The ratio of the Bayes factors from these cases can the  be used to compute the glitch/gravitational wave
Bayes factor.

\begin{figure}[ht]\label{fig:evidence}
\centerline{\includegraphics[scale=0.5]{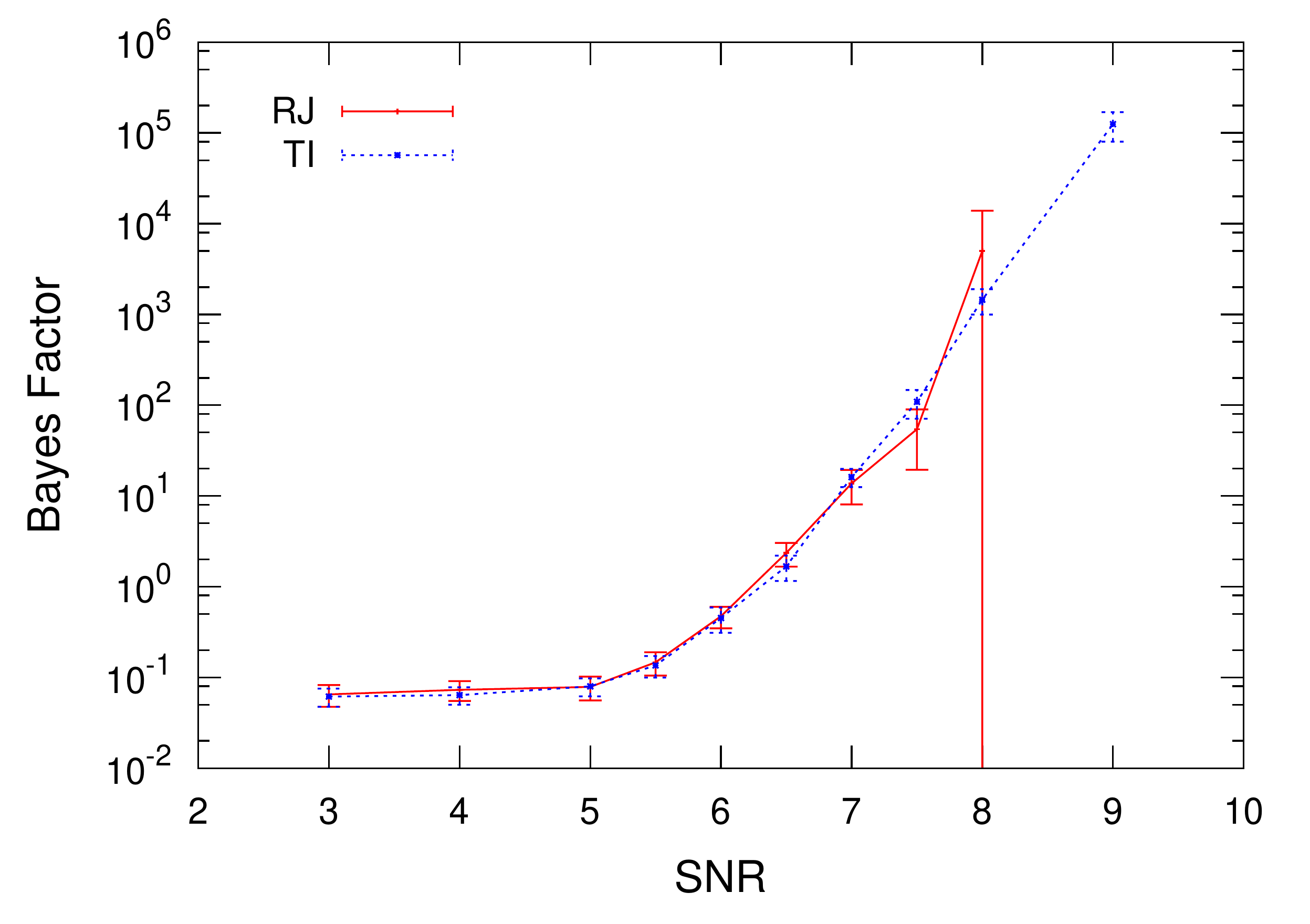}}
\caption{Bayes factors between the Gaussian-noise plus glitch model versus the Gaussian-noise model for simulated single Sine-Gaussians with $Q=12.7$ and $f_0 = 225$ Hz in simulated Gaussian noise. The
solid (red) line is computed from RJMCMC transitions, while the dashed (blue) line is computed via thermodynamic integration. Our implementation of the RJMCMC approach is unable to compute Bayes factors
in excess of $\sim 10^3$, which explains the termination of the solid (red) line at  ${\rm SNR}=8$ and the large error bar on this final value. }
\end{figure}

However, even the pairwise model comparisons become challenging for loud events as the chains spend very little time in the Gaussian-noise model, and make few transitions between
the models, which results in large statistical errors in the Bayes factors. To overcome this problem we adopted a hybrid approach that uses a RJMCMC algorithm to marginalize over the parameters in
each composite model (including the number of spline points, wavelets {\it etc}), and Thermodynamic Integration~\cite{Littenberg:2009bm} to compute the evidence for the composite model via
an integral of the average log likelihood as a function of ``inverse temperature'' $\beta = 1/T$:
\begin{equation}\label{TI}
\ln p(s\vert M) = \int_0^1 d\beta \, {\rm E}_\beta[\ln p(s\vert \vec{\lambda}, M)]
\end{equation}
Here the expectation value for the log likelihood for a chain with inverse temperature $\beta$, ${\rm E}_\beta[\ln p(s\vert \vec{\lambda}, M)]$, averages over both the model parameters and the model
dimension of the composite model $M$. To compute the integral (\ref{TI}) we first change variables to $\log\beta$ and discretize using a uniform spacing in $\log\beta$. Since we
already use parallel tempering to promote mixing of the Markov chains, we have a ready made temperature ladder with which to compute the evidence.

Figure 7 compares Bayes factors computed using thermodynamic integration and from RJMCMC transitions for a simple test case with a simulated Sine-Gaussian ``glitch'' in simulated Gaussian noise.
The two methods agree very well for low SNR glitches, but beyond ${\rm SNR}=7.5$ our implementation of the RJMCMC technique fails as the Markov chains rarely transition away from the glitch model.
It is possible to push the RJMCMC technique to higher SNR and higher Bayes factors by introducing pseudo priors on the models so that the chains spend roughly equal time in each model~\cite{RJp}, but even then
it is difficult to find proposals that allow for frequent transitions between the models. 

\subsection{Evidence Error Estimation}

When using Bayesian evidence to perform model selection, such as deciding if a certain non-Gaussian feature in the data is an instrumental artifact or a gravitational wave signal, it is important to
have reliable estimates of the evidence {\em and} the error in the evidence estimate. One standard technique is to repeat the analysis multiple times using different random number seeds and compute
the variance of the evidence estimates, but this can be extremely costly, especially if we demand better than $\sim 30\%$ accuracy in the error estimates. A much cheaper alternative is to develop
internal error estimates that can be computed from a single run. The Nested Sampling technique~\cite{skilling} comes with an internal error estimate, but we were unable to find similar estimates for
the RJMCMC and thermodynamic integration techniques in the literature, so we derive them here.

\begin{figure}[ht]\label{fig:evidenceerror}
\centerline{\includegraphics[scale=0.5]{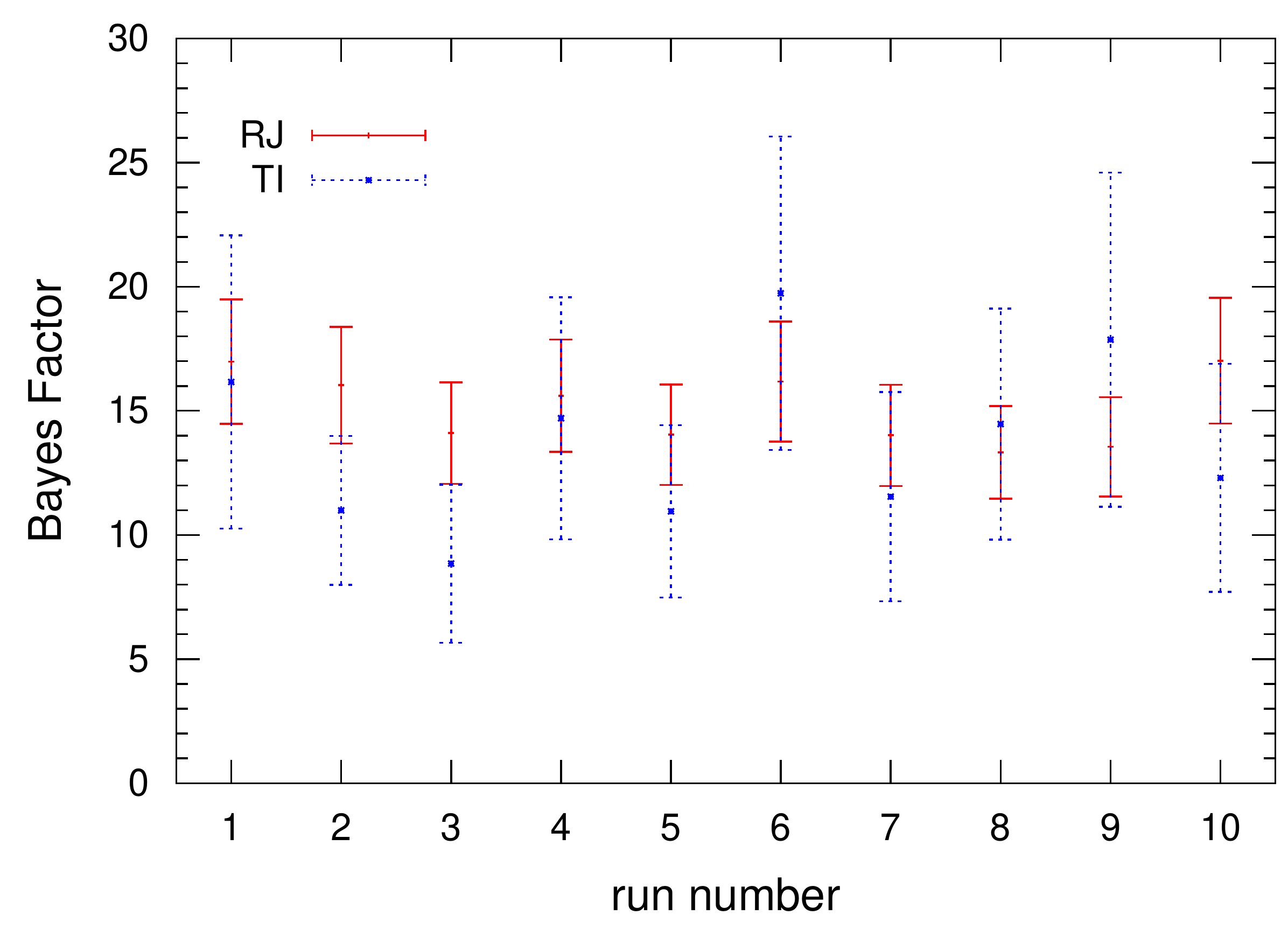}}
\caption{A scatter plot showing Bayes Factors from multiple runs on simulated data with a ${\rm SNR}=7$ Sine-Gaussian ``glitch''.  The solid (red) line is for the RJMCMC method, while the dashed (blue) line is
for the thermodynamic integration method. The error bars are the ``2-sigma'' ranges computed using our within-run error estimates.}
\end{figure}

\subsubsection{RJMCMC Evidence Error Estimate}

Here we derive the statistical error in the Bayes factor computed by the RJMCMC approach. The extension to cases with three or more models is straightforward. If we focus on the
model parameter indicator, so effectively marginalizing over the parameters of each model, the Markov chain from a RJMCMC study can be described by the two-state transition matrix~\cite{willfarr}
\begin{equation}
T =  \left( \begin{array}{cc}
1-p_{12} & p_{12}\\
p_{21} & 1- p_{21}
\end{array} \right)
\end{equation}
where $p_{ij}$ is the transition probability between states $i \rightarrow j$. The left eigenvalues of the transition matrix are $\lambda_1 = 1$ and $\lambda_2 = 1-(p_{12}+p_{21}) < 1$. As we iterate the Markov
chain the components of the initial state that lie along the eigenvector corresponding to the smaller eigenalue $\lambda_2$ decay away, and we are left with the stationary state which is given by
the $\lambda_1$ eigenvector with components
\begin{equation}
\vec{e}_1 \rightarrow \left( \frac{p_{21}}{p_{12}+p_{21}}, \frac{p_{12}}{p_{12}+p_{21}} \right) \, .
\end{equation}
The Bayes factor $B_{12}$ is simply the relative probability for the two models in the stationary state: $B_{12} = p_{21}/p_{12}$. In practice we can only estimate the transition probabilities
from number of counts in each model $N_{ii}$ and the number of transitions between models $N_{ij}$.  The joint likelihood of observing the collection of transitions
$d=(N_{11},N_{12},N_{21},N_{22})$ is given by
\begin{equation}
p(d\vert p_{ij}) = \frac{(N_1+1)! (N_2+1)!}{N_{11}! N_{12}! N_{22}! N_{21}!}(1-p_{12})^{N_{11}}p_{12}^{N_{12}}(1-p_{21})^{N_{22}}p_{21}^{N_{21}} \, ,
\end{equation}
with $N_i = N_{i1}+N_{i2}$. Setting $\partial_{p_{ij}} p(d\vert p_{ij}) =0$ yields the maximum likelihood estimate
\begin{equation}
\hat{p}_{12}= \frac{N_{12}}{N_1}\, , \quad \hat{p}_{21} = \frac{N_{21}}{N_2} \, .
\end{equation}
The maximum likelihood estimate for the Bayes factor is given by $\hat{B}_{12} = \hat{p}_{21}/\hat{p}_{12} = N_1/N_2$ since in the large $N$ limit $N_{12}=N_{21}$.
The covariance in these estimates can be found from the inverse, $C_{\mu\nu}$, of the Fisher information matrix $\Gamma_{\mu\nu} = -\partial_\mu \partial_\nu \ln p(d\vert \mu)$. Here
we have used the shorthand $\mu=p_{ij}$. The covariance matrix has the form
\begin{equation}
C =  \left( \begin{array}{cc}
\frac{N_{11} N_{12}}{N_1^3}& 0\\
0 & \frac{N_{22} N_{21}}{N_2^3}
\end{array} \right) \, ,
\end{equation}
and using the identity
\begin{equation}
{\rm Var}\left(\frac{a}{b}\right) = \left(\frac{{\rm E}(a)}{{\rm E}(b)}\right)^2 \left( \frac{{\rm Var}(a)}{{\rm E}(a)^2}+\frac{{\rm Var}(b)}{{\rm E}(b)^2}-2\frac{{\rm Cov}(a,b)}{{\rm E}(a){\rm E}(b)}\right) \, ,
\end{equation}
it follows that
\begin{equation}\label{BFerr}
{\rm Var}(B_{12}) = B_{12}^2 \left( \frac{(N_1-N_{12})}{N_1 N_{12}} +\frac{(N_2-N_{21})}{N_2 N_{21}}  \right).
\end{equation}
We have verified the accuracy of this error estimate using extensive Monte Carlo studies of two state Markov chains. We also find very good agreement between this error estimate and the
brute-force error estimate found from performing multiple runs (see Figure 8). The one caveat is that the marginalization over the model parameters for each model needs to be efficient,
otherwise our starting assumption that the transitions can be modeled by a two state Markov chain is violated. If the within-model mixing is poor, the estimate (\ref{BFerr}) provides a lower
bound on the error. For poorly mixed chains the error estimate may need to be inflated by as much as a factor of 2.

\subsubsection{Thermodynamic Integration Evidence Error Estimate}

The log evidence for model $M$ given data $s$ can be computed:
\begin{equation}\label{TIlog}
\ln p(s\vert M) = \int_{-\infty}^0 \beta \, {\rm E}_\beta[\ln p(s\vert \vec{\lambda}, M)] d\ln \beta
\end{equation}
In practice the integral is evaluated by discretizing $\ln \beta$ and estimating the expectation value from the average
log likelihood for chains running at the various inverse temperatures. Numerical error will result from
both the discretization of the integral and from the statistical error in the estimate of the average log
likelihoods.  We have developed a numerical technique for estimating both sources of error that is described below.
It is also useful to derive an analytic estimate for the statistical error that can be compared to the numerical estimate.

Defining $x= \ln \beta$, $y=\beta\, {\rm E}_\beta[\ln p(s\vert \vec{\lambda}, M)]$, and   $\Delta x_i = x_i- x_{i-1}$, we can develop a trapezoid discrectization of the integral:
\begin{equation}
 I=\int y \, d x \simeq \sum_{i = 1}^K \ \frac{1}{2} (y_i+y_{i-1})\Delta x_i
\end{equation}
Now suppose that our estimate for $y_i$ is centered at $\bar{y}_i$ and has variance $\sigma_i^2$. Let us
further assume that the errors at different temperatures are uncorrelated: $E[(y_i -\bar{y}_i)(y_j-\bar{y}_j)] = \sigma_i^2 \delta_{ij}$.
Then
\begin{equation}
 E[I] \simeq \sum_{i = 1}^K \frac{1}{2} (\bar{y}_i+\bar{y}_{i-1})\Delta x_i
\end{equation}
and
\begin{equation}\label{error}
 {\rm Var}[I]  \simeq \sum_{i =1}^K \frac{1}{4}\left( \sigma^2_i \,\Delta x_i^2 + 2 \sigma^2_{i-1} \Delta x_i\Delta x_{i-1} + \sigma_{i-1}^2 \Delta x_i^2\right) \, .
\end{equation}
The only subtle point to remember when computing this error estimate for thermodynamic integration is that the quantity $y_i$ is
found as the average of a collection of samples. Thus we need the error on the mean, which for a large number of {\em independent} samples $N$
is given by the central limit theorem as $\sigma_i^2 = {\rm Var}[y_i]/N$. Since we use a RJMCMC algorithm, our samples are correlated, and to estimate $N$
we first have to thin the chains by an amount determined by their auto-correlation length.

Depending on the spacing of the chains and the length of the runs, the discretization error may be larger or smaller than the statistical error. While it is
possible to estimate the discretization error analytically, the expression involves derivatives of the average likelihood which are hard to compute in a reliable
fashion. Instead we have developed a novel approach that simultaneously accounts for the discretization error and the statistical error. Taking as input the
estimates $\bar{y}_i$ and their variances $\sigma_i^2$, we fit a smooth curve to the data and compute the integral using the smooth curve. The curve is
defined by a cubic spline, where the number of spline control points and their location can be varied. We define a likelihood $\sim e^{-\chi^2/2}$ with
\begin{equation}
\chi^2 =  \sum_{i = 0}^K \frac{(\bar{y}_i - c_i)^2}{\sigma_i^2} \, ,
\end{equation}
where $c_i$ are the values of the cubic spline curve at $x_i$. A RJMCMC routine (adapted from the BayesLine algorithm) marginalizes over the number and location
of the spline points, and at each iteration an estimate for the integral (\ref{TIlog}) is computed using the smooth spline curve. The expectation value and variance
of these evidence estimates provide the central value and error estimate for the thermodynamic integration approach. We concede that there is an element of irony
in the fact that we use a RJMCMC algorithm to estimate error in the thermodynamic integration evidence, but the method is very robust. 

The error estimate was tested in many ways, including the multiple-run test shown in Figure 8, and tests on known functions. For example, integrating $y=1+{\rm tanh}(x)$
in the interval $x=[-1,2]$ with 10 evenly spaced samples and errors of $\sigma=0.01$ yielded $I = 3.890496 \pm 0.006605 \pm 0.0032$ using a trapezoid integration,
where the first error is the statistical error (\ref{error}) and the second is the discretization error, and $I=3.891174 \pm 0.006745$ for the RJMCMC approach with smooth
splines. Since spline integration is much higher order than trapezoid integration, discretization error is a tiny fraction of the total error in the RJMCMC approach. The
statistical error is in very good agreement with the estimate from (\ref{error}). Note that both estimates include the exact value for the integral, $I=3.891222$, within their
error ranges. In multiple tests on different functions the RJMCMC approach consistently returned more accurate values than a simple trapezoid integration.

\subsection{Model Selection Examples}

The evidence ratio, or Bayes Factor, between competing models tells us how the prior odds ratio (betting odds) have been changed by the observed data. 
The Bayes Factor between the signal or glitch model and Gaussian noise is expected to follow the relation~\cite{Cornish:2011ys}
\begin{equation}
\ln {\rm BF} = (1-{\rm FF}^2)\frac{{\rm SNR}^2}{2}+\Delta \ln {\cal O} \, ,
\end{equation}
where ${\rm FF}$ is the fitting factor, which is found by maximizing the match (\ref{match}), and ${\cal O}$ is the Occam factor - the ratio of the prior to
posterior volume, the latter being defined as the region where some large fraction of the posterior weight is concentrated. To leading order, the term $\Delta \ln {\cal O}$
is expected to scale as $\Delta N \log({\rm FF})$, where $\Delta N$ is the difference in the number of parameters between the two models~\cite{DelPozzo:2014cla}. Empirically we find that
the {\em BayesWave} fitting factors scale roughly as $1-{\rm FF} \sim 1/{\rm SNR}$, so we expect the log Bayes factor to scale linearly with the signal-to-noise ratio. Figure 9 shows the Bayes factor
between the signal model and the Gaussian noise models for a simulated non-spinning black hole merger with masses $m_1 = m_2= 20 M_\odot$, with the expected linear scaling in SNR for
${\rm SNR} > 18$.

\begin{figure}[ht]\label{fig:BFIMR}
\centerline{\includegraphics[scale=0.5]{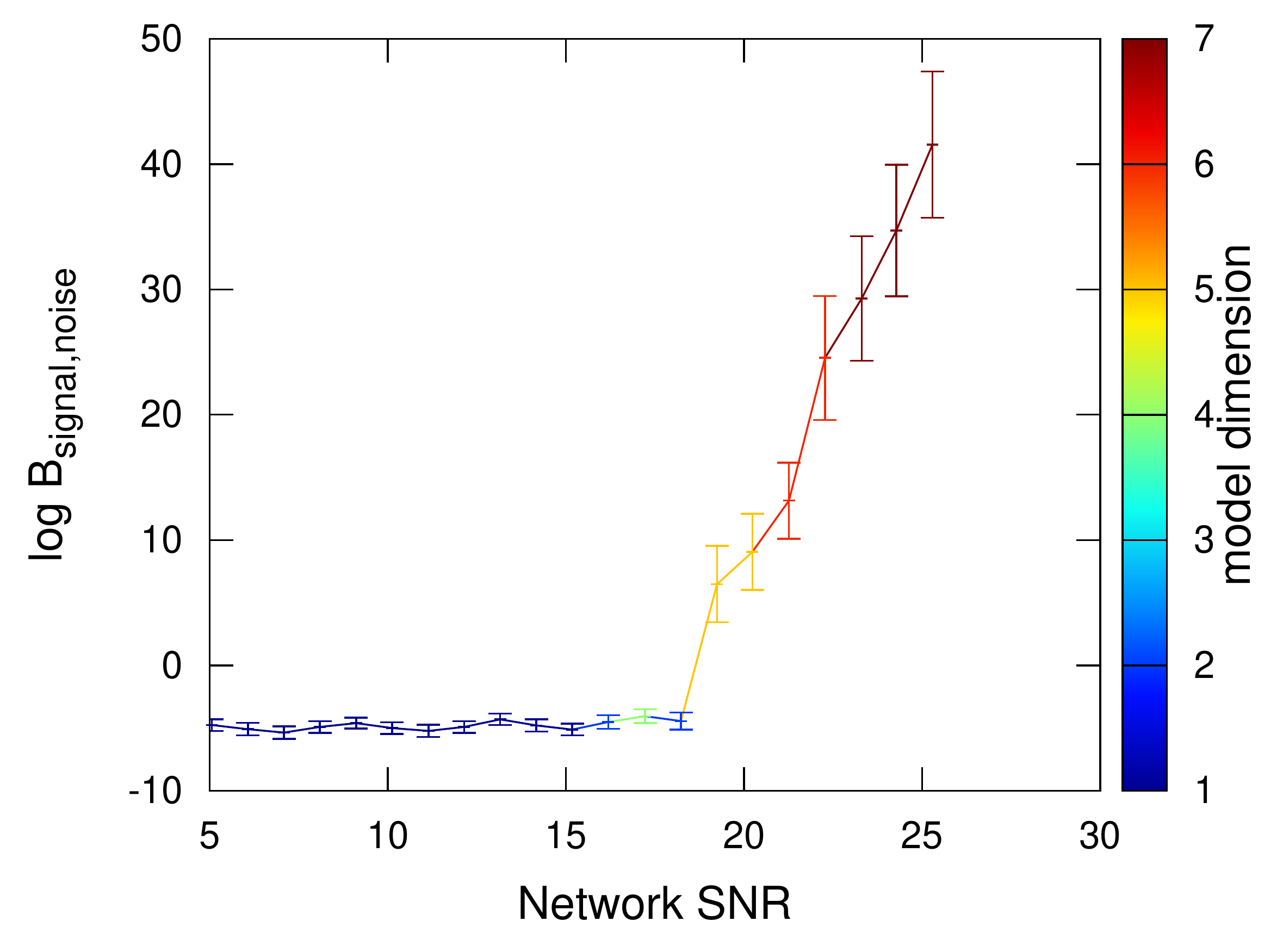}}
\caption{Bayes Factors versus SNR for a simulated black hole inspiral-merger-ringdown signal in simulated noise for the H1L1 detector network. The colors label the Maximum a Posterior number of wavelets used by
the signal model}
\end{figure}

Distinguishing non-Gaussian features in the data, be they instrument glitches or gravitational waves, from the Gaussian noise model is relatively easy. Far more challenging is
distinguishing glitches from gravitational waves. While it is extremely rare that glitches occur in every detector in the network at the same time, at the same frequency, and with a similar morphology,
the possibility exists, as demonstrated by the near coherent triggers found in time-slides of the LIGO/Virgo data~\cite{Abbott:2009zi, Abadie:2010mt}. In the absence of waveform templates, {\em BayesWave} uses similar
models for glitches and gravitational waves. Both are built from a sum of Morlet-Gabor wavelets, the only difference being that the signal model is defined at the Geocenter, and the
signals in each detector are found by projecting onto the network, while the glitch model uses independent sets of wavelets in each detector. When operating on data containing a gravitational wave
signal, both models are able to produce similar fitting factors. The signal model achieves a slightly higher fitting factor since both the match and fitting factor increase with SNR, and
the network SNR always exceeds the SNR in any one instrument. In most instances, the signal model has a lower dimensionality, and pays a lower Occam penalty as a result. As the number of
wavelets needed to reconstruct the waveform increases, or as the number of detectors in the network is increased, the signal model gains a decided advantage over the glitch model.
The worst case scenario for the signal model is when there are just two detectors, the signal has a sky location and polarization that puts most of the SNR in a single detector, and a large fraction of
the signal power can be captured by a single wavelet. In that case the glitch model can achieve a good fitting factor using 5 parameters, while the signal model is forced to use 9 parameters.
{\em BayesWave} invariably classifies these signals as glitches. Figure 10 shows signal/glitch and signal/Gaussian noise Bayes factors for a simulated Sine-Gaussian signal using different detector
combinations. The simulated signal had single detector signal-to-noise ratios of ${\rm SNR}_{\rm H1} = 12.3$,  ${\rm SNR}_{\rm L1} =9.2$,  ${\rm SNR}_{\rm V} =9.6$. In each case, there is clear evidence for
a non-Gaussian feature in the data. In the single detector analyses the signal and glitch models have comparable evidence, despite the fact that the signal model carries 4 additional extrinsic parameters.
But the signal model pays no penalty for the extra parameters since they are completely unconstrained by data from a single detector and the posterior distribution follows the prior distribution.
The signal model is heavily favored over the glitch model in all three two-detector network combinations, and the evidence for the signal model is larger still for the full three-detector network.

\begin{figure}[ht]\label{fig:BFSG}
\centerline{\includegraphics[scale=0.5]{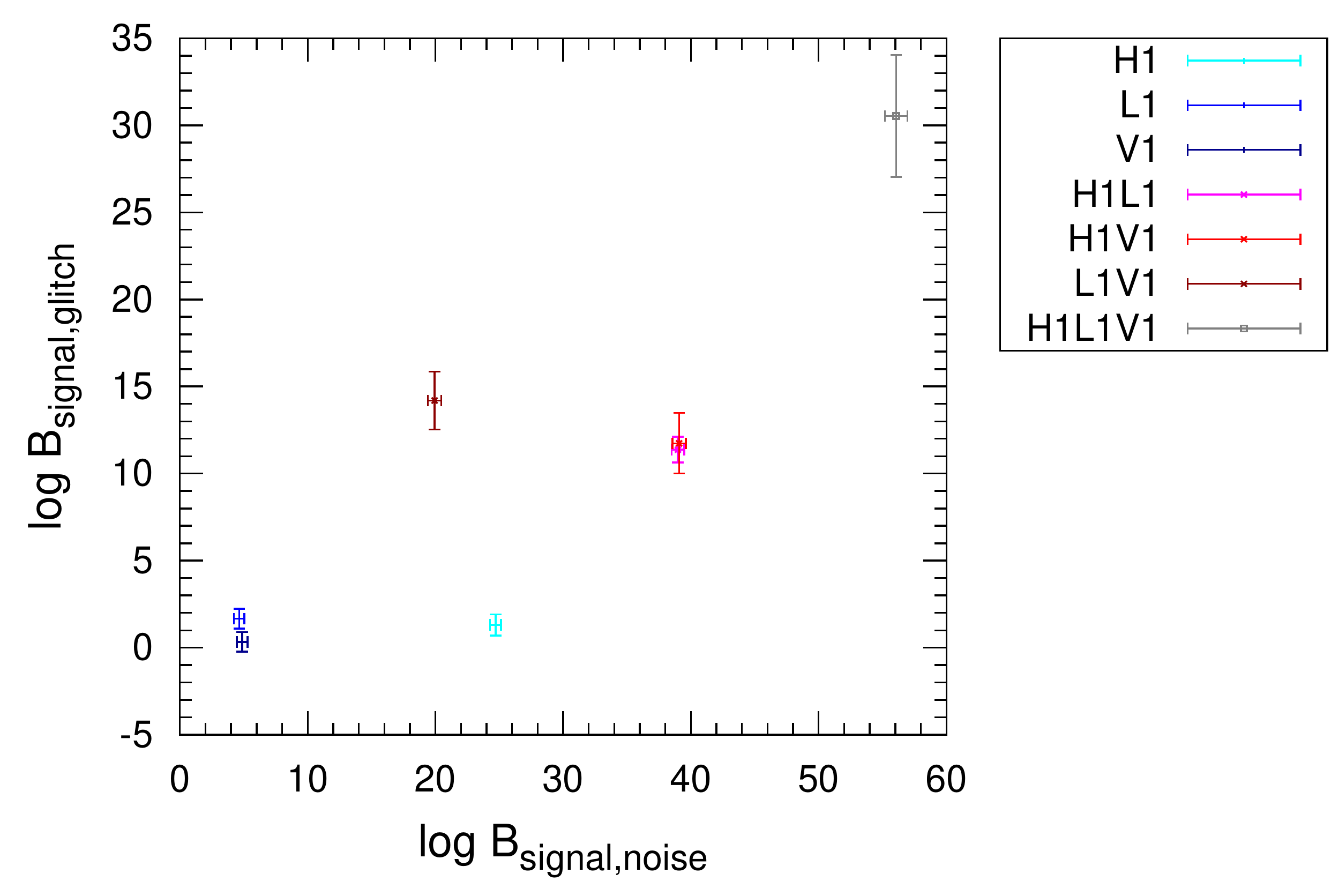}}
\caption{Bayes Factors between signal/noise and signal/glitch for simulated Sine-Gaussian signals for different detector combinations}
\end{figure}

\section{Parameter Estimation}

\begin{figure}[ht]\label{fig:waveform}
\centerline{\includegraphics[scale=0.8]{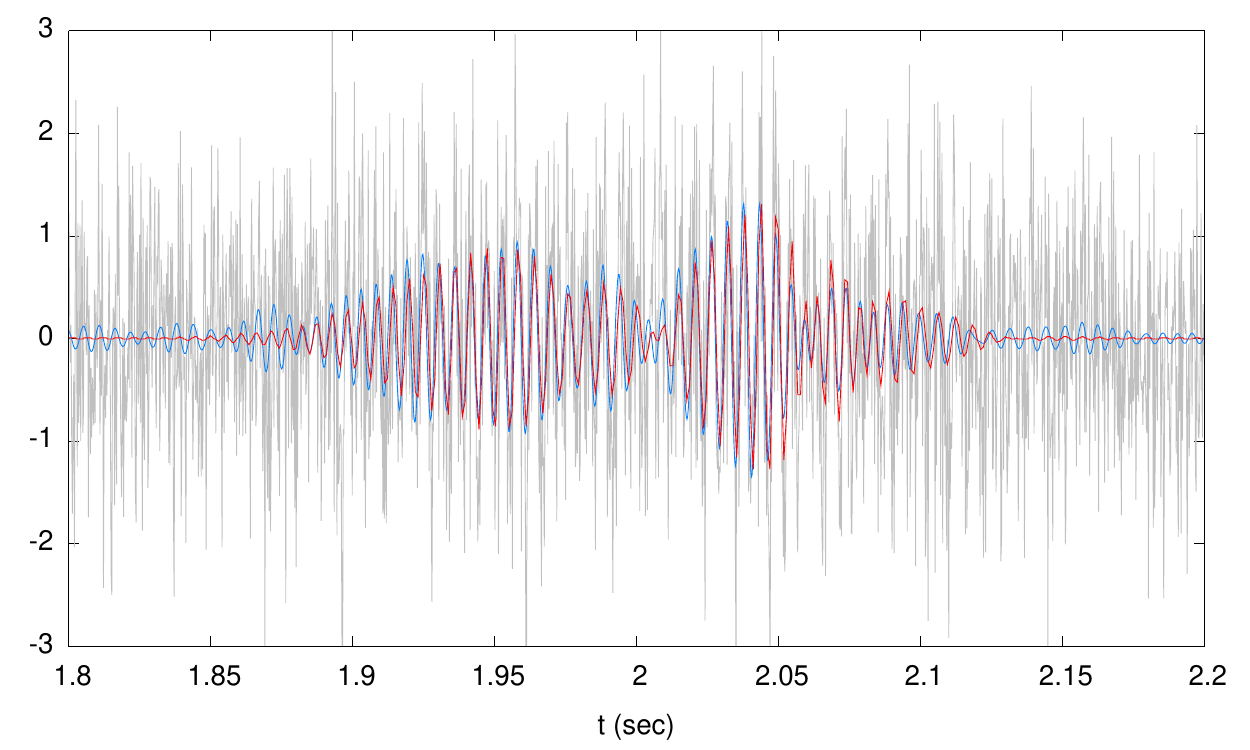}}
\centerline{\includegraphics[scale=1.0]{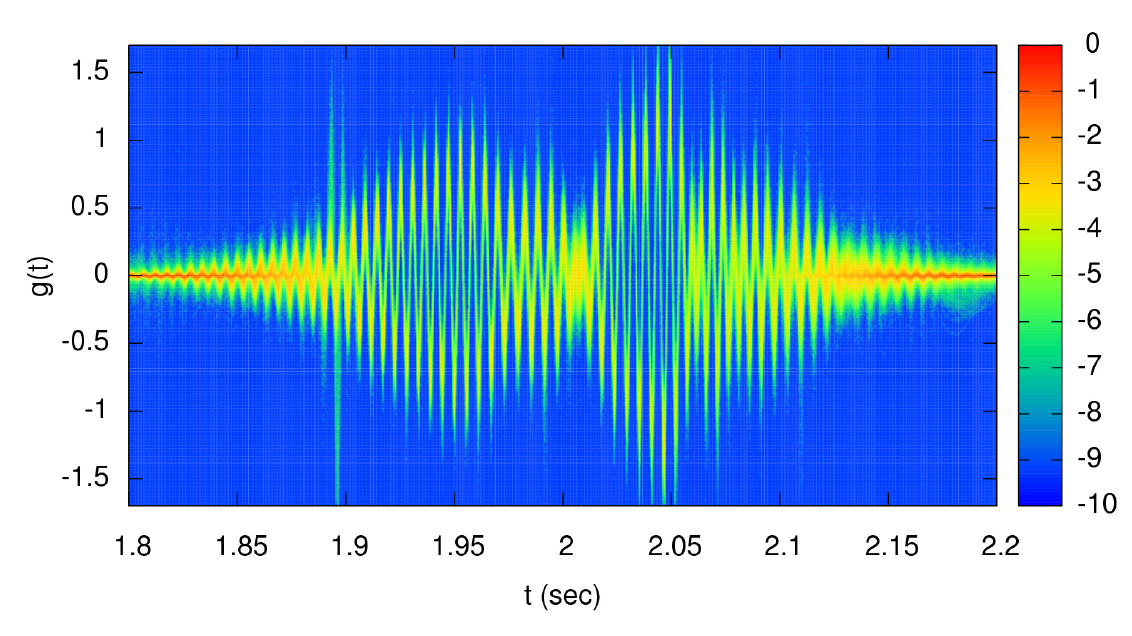}}
\caption{The upper panel shows a stretch of simulated whitened data $s(t)$ in grey. The simulation includes Gaussian noise and a ${\rm SNR=15}$ Gaussian-enveloped white-noise burst shown in blue. The red
line in the upper panel shows the median glitch model $g(t)$ reconstructed by {\em BayesWave}. The lower panel shows the posterior distribution for the glitch model, where the colors correspond to the
log posterior density $\ln p(g(t)\vert s)$.}
\end{figure}

Following the pioneering work of Christensen and Meyer~\cite{Christensen:1998gf}, the application of Bayesian inference and parameter estimation to gravitational wave astronomy~\cite{Cornish:2005qw, Cornish:2006ry,
Key:2008tt, vanderSluys:2009bf, Raymond:2009cv, Nissanke:2009kt, Veitch:2012df, Rodriguez:2013oaa, Veitch:2014wba} has mostly been focused on fully-modeled
gravitational wave signals described by waveform templates $h(\vec{\lambda})$ that depend on some collection of physical parameters $\vec{\lambda}$. Very recently Bayesian inference has been applied to
partially modeled sources such as supernovae~\cite{Rover:2009ia, Logue:2012zw,Edwards:2014uya} and Neutron star r-modes~\cite{Coughlin:2014jea}, but parameter estimation and model selection for completely
un-modeled signals has not been tackled until now (Bayesian inference for sky localization of burst signals has been considered using single sine-Gaussian waveforms~\cite{Essick:2014wwa}).

When templates are available the posterior distribution
for the gravitational wave signal $p(h\vert s)$ gets mapped to posterior distributions for the model parameters $p(\vec{\lambda}\vert s)$, allowing for a straightforward interpretation of the results, such as
producing credible intervals for the masses and spins of a binary system. But what does ``parameter estimation'' mean in the context of glitches and un-modeled gravitational wave signals? {\em BayesWave} provides posterior distributions for the wavelet parameters used to reconstruct the signals and glitches, but these have limited physical meaning. {\em BayesWave} also provides posterior distributions for the reconstructed signals $h(t)$
and glitches $g(t)$, along with posterior distributions for the sky location and polarization of a gravitational wave source. The posterior distributions for the signal and glitch waveforms contain the complete information
about the waveforms and their physical properties, but in a form that can be hard to interpret. Figure 11 shows a stretch of simulated whitened data containing a Gaussian-enveloped white-noise burst. The upper panel
shows the simulated signal and noise and the median glitch waveform reconstruction. The lower panel shows the posterior distribution for the glitch model, where the colors correspond to the
log posterior density $\ln p(g(t)\vert s)$. While providing a useful visualization, plots of the waveform posteriors do not provide a good quantitative understanding of the signal.

\begin{figure}[ht]\label{fig:PEhist}
\begin{center}
\begin{tabular}{c}
\includegraphics[scale=0.8]{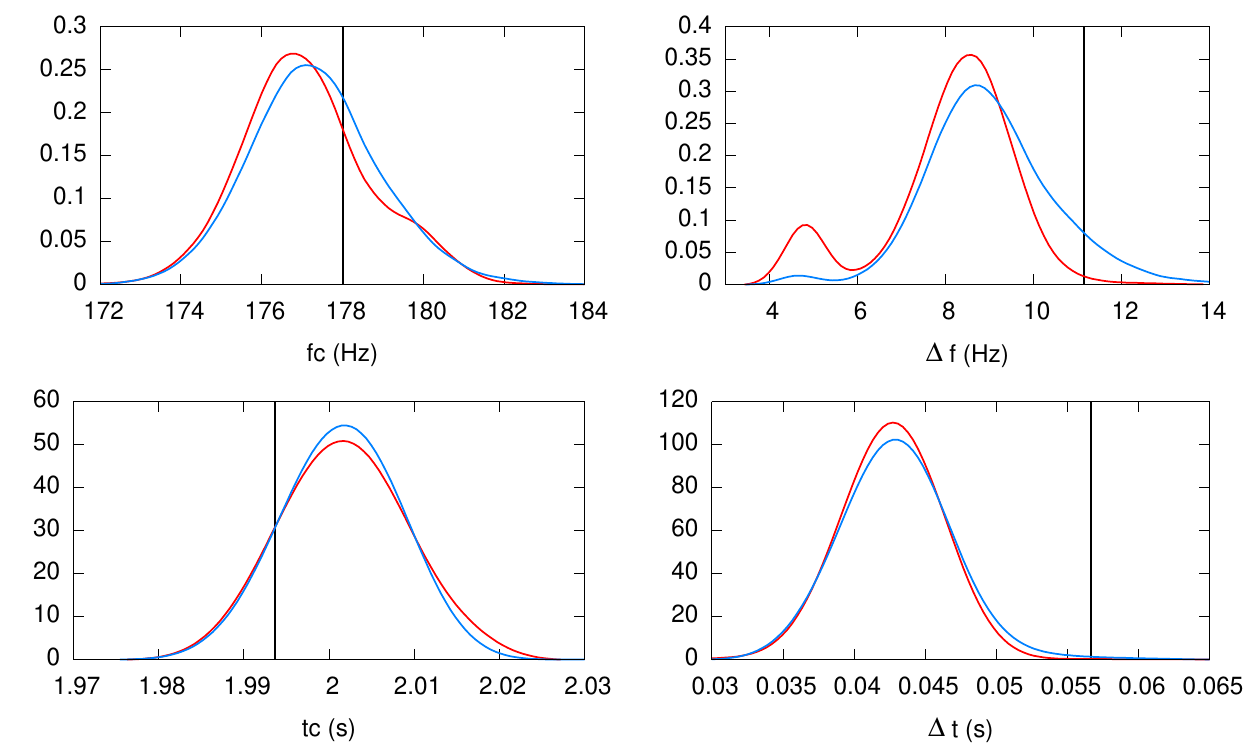} \\ \\
\hline
\\
\includegraphics[scale=0.8]{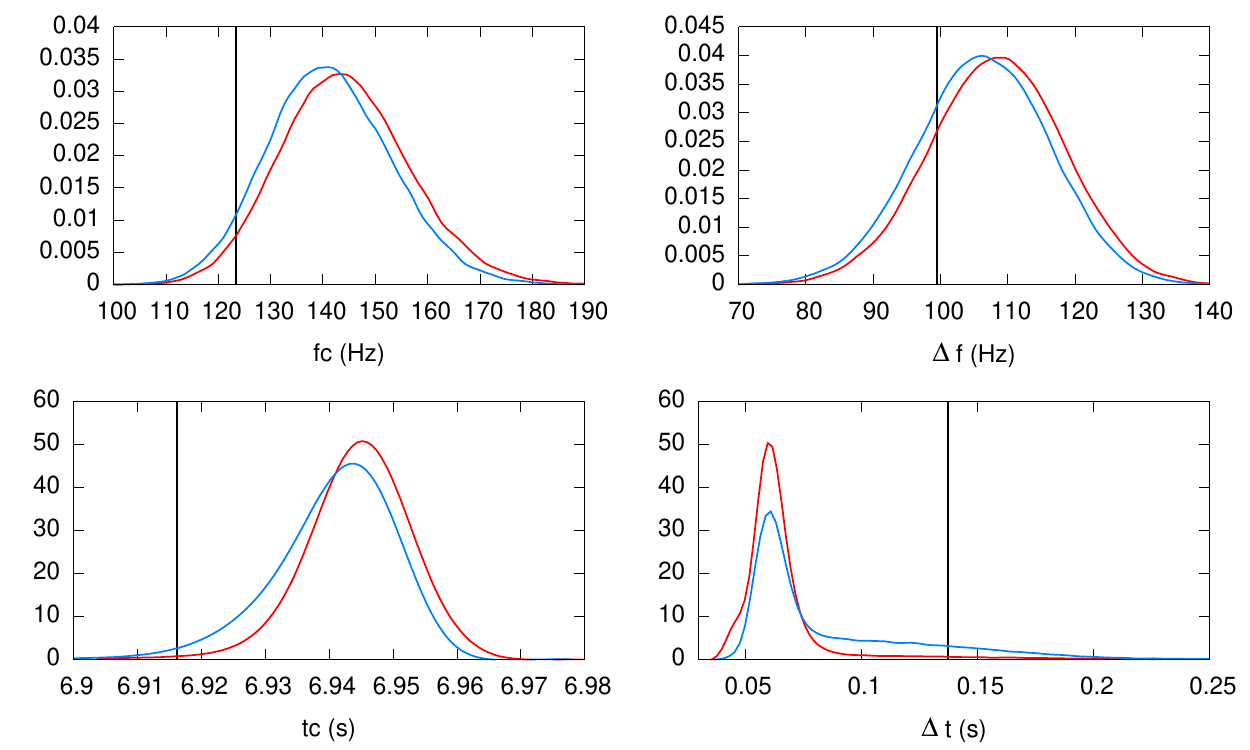}
\end{tabular}
\end{center}
\caption{The upper set of panels show posterior distributions for the central time $t_c$, central frequency $f_c$, duration $\Delta t$ and bandwidth $\Delta f$ for a ${\rm SNR=15}$ Gaussian enveloped white noise burst. The
blue lines are with the proximity prior and the red lines are with a uniform time-frequency prior. The black vertical lines indicate the values for these quantities computed with the waveform
used to generate the data. The lower set of panels show the same quantities, but this time for a ${\rm SNR=15}$ black hole merger. The parameters for the signals are the same as those used
in Figures 5, 6.}
\end{figure}

At each step in the Markov chain, {\em BayesWave} can produce the reconstructed waveforms as a function of time, or frequency, or time-frequency. The waveforms
can either be whitened, which is what the likelihood actually ``sees'', or un-whitened, which corresponds to the physical signal. While the un-whitened reconstructions would seem to be the most useful, they
tend to be very noisy as the components of the reconstruction with frequencies outside of the region where the instruments are most sensitive are poorly constrained. Using the reconstructed waveforms it
is possible to compute any physical quantity of interest, such as the energy flux, rise and decay times, duration {\it etc} at each step of the chain, and produce probability distributions for these quantities.
The choice of summary quantities to extract from the waveforms is virtually unlimited, but they will have varying utility. The most useful choices will allow us to connect the reconstructed waveforms to
astrophysical models for the signals, or for glitches, to physical models of the detector. It may be that higher dimensional quantities are needed, such as posterior distributions for Bezier curves~\cite{bezier} that
track the development of the signal in time-frequency. To illustrate the process of producing posterior distributions for parameters that summarize the physical content of the waveforms, we can use the
``waveform moments'' proposed by Patrick Sutton. The idea is to use the unit normalized densities
\begin{equation}
\rho(t) = \frac{h_+^2(t)+h_\times^2(t)}{h_{rss}^2}, \quad \rho(f) = \frac{|\tilde{h}_+(f)|^2+|\tilde{h}_\times(f)|^2}{h_{rss}^2} \, ,
\end{equation}
where 
\begin{equation}
h_{rss}^2 = \int_{-\infty}^{\infty} (h_+^2(t)+h_\times^2(t))\, dt = \int_{0}^\infty  (|\tilde{h}_+(f)|^2+|\tilde{h}_\times(f)|^2) \, df \, ,
\end{equation}
to define moments:
\begin{equation}
\langle t^n \rangle = \int_{-\infty}^{\infty}  t^n \rho(t) \, dt, \quad \langle f^n \rangle = \int_0^{\infty}  f^n \rho(f) \, df \, .
\end{equation}
The central time $t_c = \langle t \rangle$ and central frequency $f_c = \langle f \rangle$  are given by the first moments, while the duration and bandwidth are defined in terms of the variance:
\begin{equation}
\Delta t = (\langle t^2 \rangle - \langle t \rangle^2)^{1/2}, \quad \Delta f = (\langle f^2 \rangle - \langle f \rangle^2)^{1/2}\, .
\end{equation}
These quantities can be computed for both the whitened and un-whitened waveforms. Similar expressions can be defined for glitches.

Figure 12 shows posterior distributions for the central time, central frequency, duration and bandwidth (computed using whitened waveforms) for a Gaussian enveloped white
noise burst and a black hole merger using a uniform
time-frequency prior and the proximity prior. In each case, the proximity prior improves the parameter estimation, which is consistent with the improvements seen in the fitting factors for
these systems shown in Figures 5 and 6. These examples illustrate a general trend that we have seen toward systematic biases in the recovery of certain parameters, most notably the
duration of the signal. In general we find that {\em BayesWave} under-estimates the duration and energy content of the signals. These biases are to be expected given that the match
between the recovered and simulated signals is always less than unity. It may be possible to compensate for this bias since the cause is known. On the other hand, the waveform moments are
particularly susceptible to such biases, and we are probably better off using more robust summary statistics, such as defining the duration and bandwidth as the intervals containing $90\%$ of the
energy.

\section{Applications using LIGO/Virgo data}

In addition to tests on simulated data, like those shown in the preceding sections, {\em BayesWave} has been extensively tested on data from the S5/6 LIGO and VSR2/3 Virgo science runs.
These studies will be presented in full elsewhere, but as a prelude, we show a few highlights from these studies here. For reference, the runs use a time frequency volume with duration 4 s,
and frequency range $[16,512]$ Hz. Typical wall-clock runs times to complete $2 \times 10^6$ iterations for all three models (Gaussian noise, glitch and signal model) was 24 hours on
one processor core. 

\begin{figure}[ht]\label{fig:gauss}
\centerline{\includegraphics[scale=0.5]{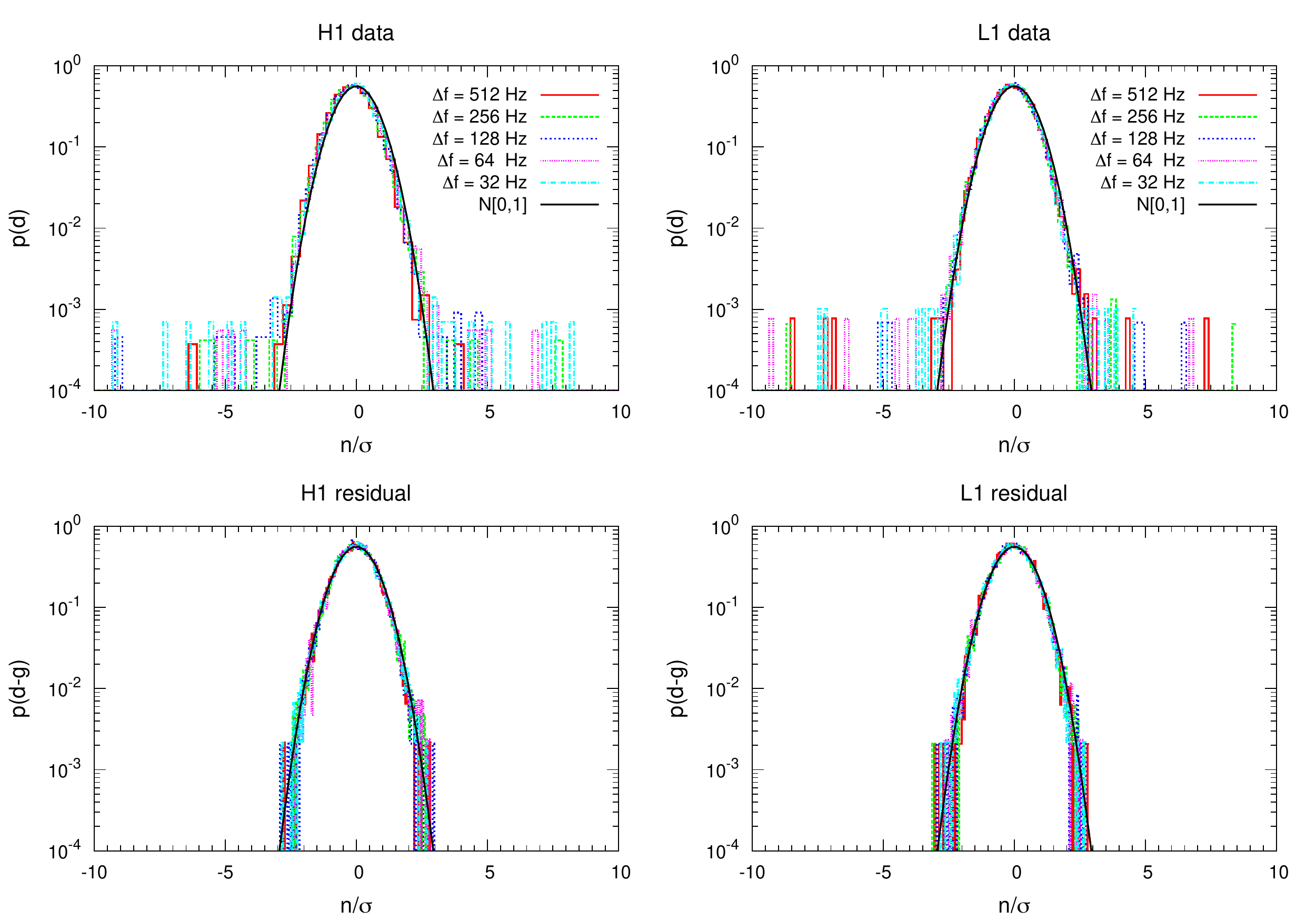}}
\caption{Histograms of wavelet amplitudes for whitened Hanford and Livingston LIGO data. The different colors in the histograms correspond to different frequency resolutions in a discrete
binary Meyer wavelet decomposition. The black line is for a reference $N(0,1)$ Gaussian distribution.
The non-Gaussian features in the data are more apparent in some resolution levels than others. The upper panels are for stretches of moderately
glitchy data prior to the {\em BayesWave} analysis, while the lower panels show the distributions after {\em BayesWave} glitch regression.
The non-Gaussian outliers seen in the raw data are absent in the {\em BayesWave} residuals.}
\end{figure}

One application of {\em BayesWave}/{\em BayesLine} is to use the noise spectrum and glitch models in concert with standard gravitational wave template based searches and parameter estimation studies
in the hope that removing glitches from the data will reduce the number of background events and lead to smaller biases in the extraction of the signal parameters when a detection is made. To illustrate the
potential of this approach, we took a stretch of moderately glitchy data from the LIGO S6 science run and produced whitened scalograms with and without glitch fitting. Histograms of the wavelet amplitudes
at various discrete wavelet resolutions are shown in Figure 13 (here we used the binary Meyer wavelet transformation developed by S. Klimenko~\cite{Klimenko:2004qh}). The discrete wavelet transform
has pixels with bandwidth $\Delta f$, duration $\Delta t$ and area $\Delta f \Delta t =1$. The histograms are shown for a variety of resolutions $\Delta f$ as glitches come with a variety of morphologies
that are picked up more clearly in some resolutions than others~\cite{Sturani:2007tc}. Note that the {\em BayesWave} residuals, with glitches regressed, are Gaussian distributed at {\em all} resolutions.

\begin{figure}[ht]\label{fig:cWB}
\centerline{\includegraphics[scale=0.5]{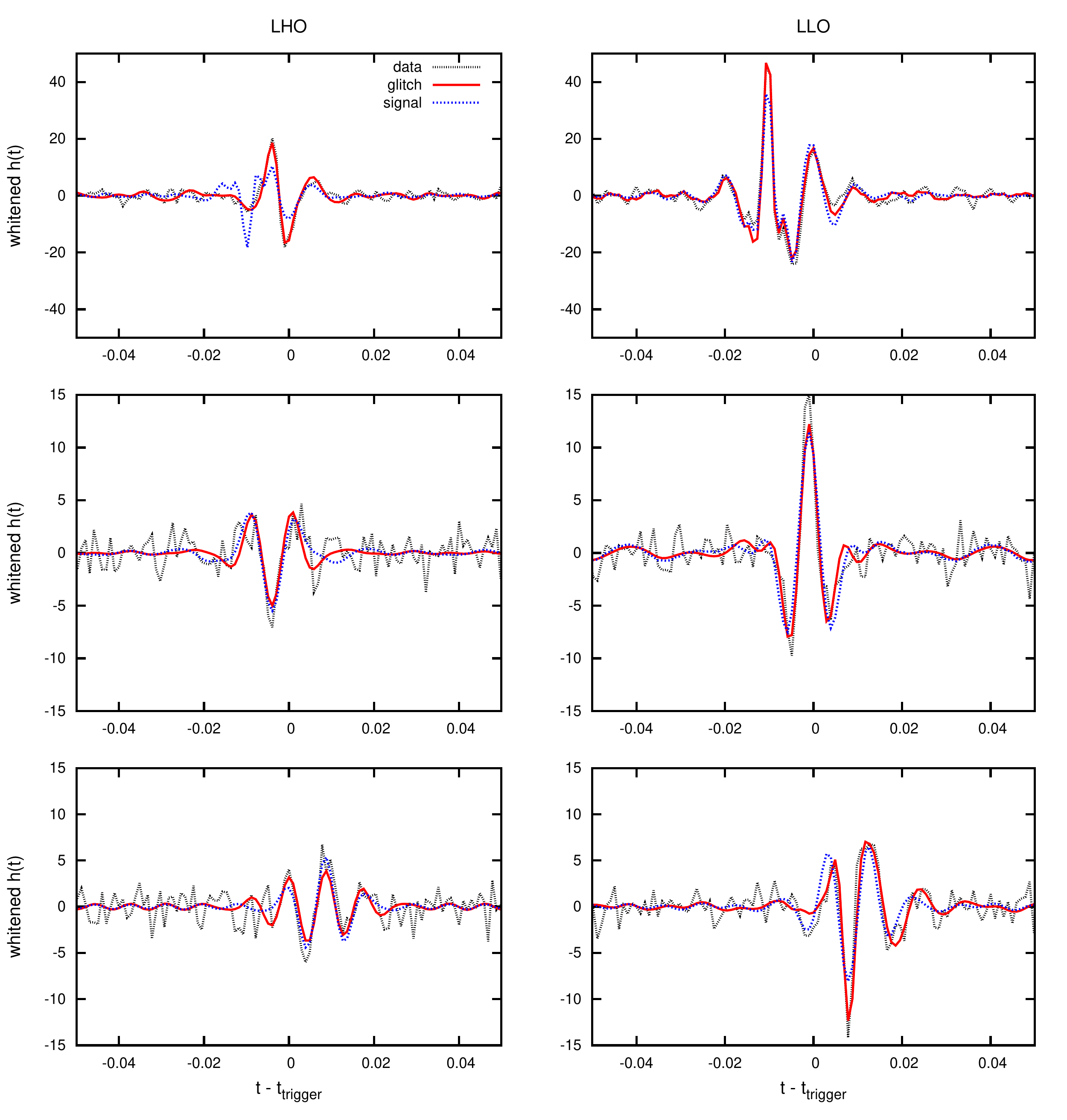}}
\caption{Snapshots of the signal and glitch model waveform reconstructions for three of the most significant events found in time-slides by the coherent Wave Burst analysis running on
the H1-L1 detector network during LIGO science run S6d.}
\end{figure}

In addition to performing parameter estimation for signals and glitches, {\em BayesWave} can be used to help decide if candidate events are gravitational wave signals or (un)lucky coincidences between
noise events in different detectors. The main burst search algorithm used by the LIGO/Virgo collaborations, coherent WaveBurst~\cite{cWB}, performs a constrained maximum likelihood reconstruction of
gravitational wave signals, supplemented by additional cuts on the ``network correlation coefficient'', the ``coherent network amplitude'' and the ``network energy disbalance'' (see equations (E23), (E27) and (E21) of
Ref.~\cite{Abbott:2009zi}), which are designed
to reject accidental coincidences from glitches. These cuts are tuned using simulated signals and time-slides of the data, the former to measure the impact on the false dismissal rate, the latter to measure the
the impact on the false alarm rate. The most significant events found in time-slides after cuts and data quality vetoes have been applied determine the overall sensitivity level for the coherent WaveBurst search
(note that cWB did not assume that the signals are elliptically polarized).
Figure 14 shows examples of the waveform reconstruction for the 4 km LIGO detectors at Hanford and Livingston taken from randomly chosen iterations of the Markov chains. In each case the signal
and glitch reconstructions are very similar, which is why these spurious coincidences masqueraded as signals in the coherent WaveBurst analysis. Going from the top panel to the bottom panel, coherent WaveBurst
reported network signal-to-noise ratios and ``network correlation coefficients'': Upper -- ${\rm SNR} = 64.4$, $cc = 0.84$; Middle -- ${\rm SNR} = 20.4$, $cc = 0.93$; Lower -- ${\rm SNR} = 18.7$, $cc = 0.88$.
The {\em BayesWave} algorithm does not compute quite the same quantities, so instead we quote the maximum signal-to-noise ratios and Bayes Factors between the signal, glitch and noise models:
Upper -- ${\rm SNR}_{\rm max} = 88.7$, $\ln {\rm BF_{s/n}} = 1872.9 \pm 0.5$, $\ln {\rm BF_{g/s}} = 24.5 \pm 3$; Middle --
${\rm SNR}_{\rm max} = 19.5$,  $\ln {\rm BF_{s/n}} = 63.8 \pm 0.5$, $\ln {\rm BF_{g/s}} =  -6.9 \pm 2$; Lower -- ${\rm SNR}_{\rm max} = 18.9$, $\ln {\rm BF_{s/n}} =  53.7\pm 0.8$, $\ln {\rm BF_{g/s}} =  -2.3 \pm 3$.
In each case {\em BayesWave} found overwhelming evidence for a non-Gaussian feature in the data.
For the event in the upper panel the evidence decisively favored the glitch model over the signal model, while for the event in the middle panel the evidence slightly favored the signal model over the glitch model.
For the event in the lower panel the evidence is inconclusive. The fact that {\em BayesWave} found the event in the middle panel to be most signal-like is consistent with this event having the highest network correlation
coefficient in the coherent WaveBurst analysis - sometimes sufficiently similar coincident glitches in two detectors can fool any Burst search.

\begin{figure}[ht]\label{fig:sky}
\centerline{\includegraphics[scale=0.6]{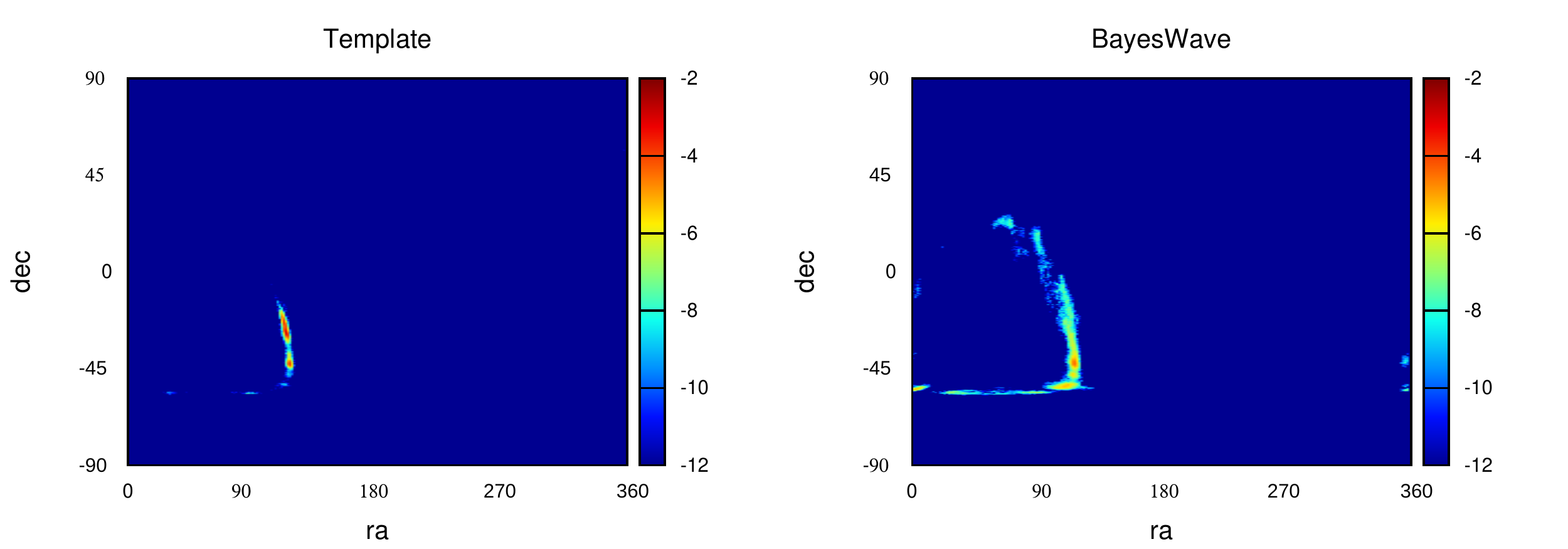}}
\caption{Sky maps for the ``Big Dog'' blind injection. The panel on the left uses waveform templates, while the panel on the right uses wavelets.}
\end{figure}

As a final example, we show in Figure 15 sky maps for a notable blind (artificial) signal injection made on September 16, 2010~\cite{bigdog}, that has variously been dubbed ``gravitational wave100916'' for the date of
discovery~\cite{Colaboration:2011np} or ``the Big Dog'', as the initial sky reconstruction placed the event in the direction of Canis Major.
The panel on the left shows the log posterior density derived using inspiral templates, while the panel on
the right used the {\em BayesWave} wavelet model. The two sky maps are broadly consistent. The {\em BayesWave} sky map is more diffuse than the template derived sky map, as is to be expected
since the wavelet model captures less of the signal power than the template model. 

\section{Conclusion and Outlook}

We have presented a new approach to gravitational wave data analysis that puts as much emphasis on modeling the instrument noise as it does on modeling the signals. Our approach follows the
motto ``model everything and let the data sort it out'', where crucially, the data gets to decide on the complexity of the model. The {\em BayesWave} algorithm provides a flexible framework for
analyzing gravitational wave data. The signal and glitch models can be changed by modifying the priors. Targeted analyses can easily be developed for particular classes of signals that build in
even partial knowledge about the waveforms. Work is currently underway to develop targeted analyses for spinning binary mergers and high eccentricity binaries that use parameterized priors
based on the time-frequency evolution of these systems. As more detectors join the world-wide network we will be able to do away with the restriction that the signals are elliptically polarized and
independently infer the two polarization states $h_+$ and $h_\times$. 

Currently, parameter estimation for gravitational wave bursts and instrument glitches is in its infancy. There is much room for improvement in the development of quantities that meaningfully summarize the
physical characteristics of the reconstructed waveforms and relate these to models for the astrophysical or electro-mechanical processes that gave rise to the signals or glitches.

\section{Acknowledgments}
We have enjoyed many productive discussions with Will Farr, Vicky Kalogera, Jonah Kanner and Reed Essick.
We have also benefitted from discussions with Sergei Klimenko and Chris Pankow on the inner workings of the coherent WaveBurst algorithm.
We appreciated feedback on the draft from Laura Sampson, Tiffany Summerscales and Jonathan Gair. We a grateful to the LIGO and Virgo
collaborations for making available the data used in Section 8 of the paper.
NJC appreciates the support of NSF Award PHY-1306702.  TBL acknowledges the support of NSF Award PHY-1307020

\end{document}